# NMR parameters in alkali, alkaline earth and rare earth fluorides from first principle calculations

**Aymeric Sadoc, \*,[a] Monique Body,[b] Christophe Legein,[c] Mamata Biswal,[c] Franck Fayon,[d] Xavier Rocquefelte,[a] and Florent Boucher[a]** [a] *Institut des Matériaux Jean Rouxel (IMN), Université de Nantes, CNRS, 2 rue de la Houssinière, BP 32229, 44322 Nantes Cedex 3, France. E-mail : Aymeric.Sadoc@cnrs-imn.fr; Fax: +33 2 40 37 39 95; Tel: +33 2 40 37 64 11*
[b] *Laboratoire de Physique de l'Etat Condensé, CNRS UMR 6087, Institut de Recherche en Ingénierie Moléculaire et Matériaux Fonctionnels (CNRS FR 2575) Université du Maine, Avenue Olivier Messiaen, 72085 Le Mans Cedex 9, France*
10 [c] *Laboratoire des Oxydes et Fluorures, CNRS UMR 6010, Institut de Recherche en Ingénierie Moléculaire et Matériaux Fonctionnels (CNRS FR 2575) Université du Maine, Avenue Olivier Messiaen, 72085 Le Mans Cedex 9, France*
[d] *Conditions Extrêmes et Matériaux: Haute Température et Irradiation, CNRS UPR 3079, 1D Avenue de la Recherche Scientifique, 45071 Orléans Cedex 2, France and Université d'Orléans, Faculté des Sciences, Avenue du Parc Floral, 45067 Orléans Cedex 2, France*

15 $^{19}$F isotropic chemical shifts for alkali, alkaline earth and rare earth of column 3 basic fluorides are measured and the corresponding isotropic chemical shieldings are calculated using the GIPAW method. When using PBE exchange correlation functional for the treatment of the cationic localized empty orbitals of $Ca^{2+}$, $Sc^{3+}$ (3d) and $La^{3+}$ (4f), a correction is needed to accurately calculate $^{19}$F chemical shieldings. We show that the correlation between experimental isotropic
20 chemical shifts and calculated isotropic chemical shieldings established for the studied compounds allows to predict $^{19}$F NMR spectra of crystalline compounds with a relatively good accuracy. In addition, we experimentally determine the quadrupolar parameters of $^{25}$Mg in $MgF_2$ and calculate the electric field gradient of $^{25}$Mg in $MgF_2$ and $^{139}$La in $LaF_3$ using both PAW and LAPW methods. The orientation of the EFG components in the crystallographic frame, provided by DFT
25 calculations, is analysed in term of electron densities. It is shown that consideration of the quadrupolar charge deformation is essential for the analysis of slightly distorted environments or highly irregular polyhedra.

## Introduction

During the last decade, the characterisation of the fluorine environment in rigid solids by nuclear magnetic resonance (NMR)
30 spectroscopy has become easier with the increase in routinely available magic angle spinning (MAS) frequency which allows an efficient averaging of the chemical shift anisotropy and dipolar interactions. As the $^{19}$F ($I = 1/2$) isotropic chemical shift ($\delta_{iso}$) is very sensitive to the environment of the fluorine atom, MAS NMR is a powerful structural tool for studying complex fluoride crystalline materials having multiple crystallographic sites. In numerous studies, the interpretation of $^{19}$F MAS NMR spectra is commonly based on the intuitive assumption that similar structural environments lead to similar $^{19}$F $\delta_{iso}$.[1-8] By comparison with
35 the $^{19}$F $\delta_{iso}$ values measured for well-known binary fluorides, the $^{19}$F resonances of a crystalline compound can be assigned to different fluorine environments. In the case of fluorine sites with different multiplicities, the relative intensities of the corresponding resonances also provide additional constraints for the assignment. Nonetheless, complete unambiguous assignment of complex $^{19}$F solid-state MAS NMR spectra often remains challenging. In such cases, two-dimensional (2D) NMR correlation experiments, which provide information about inter-atomic connectivities, can be used for line assignment purposes. In inorganic
40 crystalline fluorides, various 2D heteronuclear correlation MAS experiments (CP-MAS HETCOR,[9] TEDOR-MQMAS,[10] CP 3QMAS[11] and J-HMQC[12,13]) have been applied to several spin pairs ($^{19}$F/$^{27}$Al,[14-17] $^{19}$F/$^{23}$Na,[16,18-20] $^{19}$F/$^{31}$P[14,21] and $^{19}$F/$^{207}$Pb[22-24]) to probe heteronuclear spatial proximities. In oxyfluoride[25,26] and in fluoride materials,[17,23,24,27] the fluorine-fluorine proximities or through bond connectivities evidenced through 2D $^{19}$F double-quantum single-quantum[28] (DQ-SQ) MAS correlation experiments were also used to assign the $^{19}$F resonances. However, in the case of distinct fluorine sites having the same connectivity scheme
45 and relatively similar inter-atomic distances, these 2D correlation NMR methods do not allow a straightforward assignment of the corresponding resonances.[17,29]

An alternative approach is to correlate experimental $^{19}$F $\delta_{iso}$ values to the ones calculated from structural data, Semi-empirical model[30-35] can be used for this purpose but this requires the refinement of phenomenological parameters which are usually valid for a specific family of compounds. First-principles molecular calculations are very efficient on molecular systems but, in the case of crystalline compounds, these methods critically require the non trivial definition of cluster size to mimic the crystalline structure.[25,36-45] Moreover, for these two approaches, uncertainties on calculation results are sometimes larger than the experimentally measured $^{19}$F $\delta_{iso}$ difference between two distinct resonances, preventing an unambiguous assignment of the $^{19}$F NMR resonances.

The more elegant approach for crystalline systems consists in using the periodic boundary conditions. Two different methods can be used for this purpose, the GIPAW (Gauge Including Projector Augmented Wave) method introduced by Pickard *et al.*[46-48] that enables the calculation of the chemical shielding tensor ($\sigma$) and indirect spin-spin ($J$) coupling constant[26,49-53] and the recently implemented "converse approach"[54] that was demonstrated to be a very efficient alternative for the chemical shielding tensor calculations.[55] Two groups have recently published GIPAW calculations on fluorides. From $^{19}$F isotropic chemical shieldings ($\sigma_{iso}$) calculations for numerous compounds including alkali and alkaline earth basic fluorides, Zheng *et al.*[56] proposed a calibration curve between calculated and experimental $^{19}$F $\delta_{iso}$ values. Griffin *et al.*[26] have also calculated $^{19}$F $\sigma_{iso}$ values for several fluorides including some alkali and alkaline earth fluorides and a rare earth fluoride, $LaF_3$. However, using the same exchange/correlation functional, a significantly different calibration curve was obtained. The origin of this difference is discussed later on and arises mostly from the consideration of $LaF_3$ in the correlation reported by Griffin *et al.*[26]. One can also notice in the paper of Zheng *et al.*[56] some significant differences between experimental and calculated $\delta_{iso}$ values, especially for $CaF_2$ (30 ppm).[56]

To further investigate these problems we decided to reconsider this crucial step for providing predictive results allowing the assignment of $^{19}$F NMR resonances, *i.e.* the definition of a calibration curve for inorganic fluorides. We have thus calculated the $^{19}$F $\sigma_{iso}$ for alkali, alkaline earth and rare earth (column 3) basic fluorides using the CASTEP code.[57] To obtain reliable experimental data and avoid any reference problem (see below for more details), the experimental $^{19}$F $\delta_{iso}$ values for all the compounds under investigation have been measured again using the same reference sample ($CFCl_3$). In a first step, the fluorine pseudopotential used for the calculation of $^{19}$F $\sigma_{iso}$ has been validated. As it is classically done,[55,58,59] a molecular benchmark was used for comparing our GIPAW results to all-electrons (AE) calculations. In a second step, the correlation between the calculated $^{19}$F $\sigma_{iso}$ values for twelve binary crystalline compounds and experimental $^{19}$F $\delta_{iso}$ values is investigated. A critical problem that was already observed for Ca in oxides[60] is evidenced in fluorides: PBE-DFT (Perdew, Burke and Ernzerhof - Density Functional Theory) method[61] is deficient in describing 3d and 4f localized empty orbitals when considering NMR shielding calculations. To circumvent this problem the Ca, Sc, and La pseudopotentials have been adapted using the methodology described in [60] and a reference calibration curve is proposed. We then show that the correlation established for the studied compounds fit nicely with the calculations on others inorganic fluorides reported in [56] and [26].

For two of the twelve studied compounds ($MgF_2$ and $LaF_3$), the quadrupolar nuclei occupying the cationic site (*i.e.* $^{25}$Mg, $I = 5/2$ and $^{139}$La, $I = 7/2$) are affected by the quadrupolar interaction since the corresponding site symmetries lead to non-zero Electric Field Gradient (EFG). We have measured the $^{25}$Mg NMR parameters in $MgF_2$ which were unknown despite two recent $^{25}$Mg NMR studies of numerous compounds[62,63] and the recently determined $^{139}$La NMR parameters in $LaF_3$[64,65] are also reported. The EFG tensors of $^{25}$Mg in $MgF_2$ and $^{139}$La in $LaF_3$ calculated from AE method and projector augmented-wave (PAW) approach[66,67] using the WIEN2K[68,69] and CASTEP codes,[57] respectively, are compared to these experimental values. Finally, the orientation of the EFG tensor components in terms of site distortion and deformation of the electronic density around the cationic positions is discussed.

**Experimental and computational details**

**Solid state NMR**

Experimental conditions used to record $^{19}$F solid-state MAS NMR spectra are given as ESI. The $^{25}$Mg MAS (7 kHz) NMR spectra of $MgF_2$ were recorded at two magnetic fields of 17.6 T and 9.4 T using Avance 750 and 400 Bruker spectrometers operating at Larmor frequencies of 45.92 MHz and 24.49 MHz, respectively. A Hahn echo pulse sequence with a 5.0 µs 90° pulse (nutation frequency of 50 kHz) was employed. The inter-pulse delays were synchronized with the rotor period and $^{19}$F continuous wave decoupling was applied during signal acquisition. The recycle delays were set to 5 s and 1.5 s at 17.6 T and 9.4 T, respectively. The $^{25}$Mg chemical shift was referenced relative to an aqueous 1 M solution of MgCl$_2$. All the NMR spectra were reconstructed using the DMFit software.[70]

**Computational methods**

The GIPAW method as implemented in the CASTEP code is an efficient and accurate method for determining NMR shielding tensor in periodic systems. By combining plane-wave basis set and Ultrasoft Pseudopotential (USPP) a quite large number of atoms can be considered using periodic boundary conditions. However, the pseudopotential construction (mainly the GIPAW projectors definition) should be realized with care in order to avoid unphysical behaviour that could lead to misleading conclusions.

**Table 1** Molecules used for pseudopotential tests. Local point groups, space groups, bond lengths and angles are reported. The space group is used for calculation with periodic code.

| Molecule | Local point group | Space group | Distances/Å | | Angles/° | |
|---|---|---|---|---|---|---|
| CH$_3$F | C$_s$ | P$_{31m}$ | C-F = 1.403 | C-H = 1.099 | F-C-H = 108.7 | |
| HF | C$_{\infty v}$ | P$_{4mm}$ | H-F = 0.940 | | | |
| C$_6$F$_6$ | D$_{2h}$ | P$_{6/mmm}$ | C-F = 1.343 | C-C = 1.395 | C-C-C = 120.0 | F-C-C = 120.0 |
| CH$_2$F$_2$ | C$_{2v}$ | P$_{mm2}$ | C-F = 1.376 | C-H = 1.100 | H-C-H = 113.9 | F-C-F = 108.5 |
| CF$_4$ | T$_d$ | P$_{-4m3}$ | C-F = 1.342 | | F-C-F = 109.5 | |
| CFCl$_3$ | C$_s$ | P$_{31m}$ | C-F = 1.363 | C-Cl = 1.769 | Cl-C-Cl = 110.6 | Cl-C-F = 108.3 |
| NF$_3$ | C$_{3v}$ | P$_{31m}$ | N-F = 1.409 | | F-N-F = 101.7 | |
| F$_2$ | D$_{\infty h}$ | P$_{4/mmm}$ | F-F = 1.418 | | | |

To test the validity of the GIPAW USPP used to calculate $^{19}$F σ$_{iso}$, a molecular benchmark of eight experimentally well characterized simple molecules is used (see Table 1). They were chosen because they span on a large range of $^{19}$F σ$_{iso}$ values (about 750 ppm). Two sets of calculations are performed, the first one using AE basis sets as implemented in the Gaussian03 code[71] and the second one using USPP and the GIPAW method as implemented in the CASTEP 5.0 package. For the AE calculations, the well known GIAO (gauge invariant atomic orbitals)[72,73] and IGAIM (individual gauges for atoms in molecules)[74,75] methods are used. AE calculations are performed using four different types of basis sets from Dunning's hierarchy[76] with increasing accuracy, namely aug-cc-pCVDZ, aug-cc-pCVTZ, aug-cc-pCVQZ and aug-pCV5Z taken from the ESML basis set exchange library.[77] The USPP are generated using the on the fly generator (OTF_USPP) included in CASTEP and the following parameters for the fluorine atoms: (i) r$_{loc}$ = r$_{nloc}$ = 1.4 a.u., (ii) r$_{aug}$ = 1.0 a.u. and (ii) q$_c$ = 7.5 a.u.$^{1/2}$. Two ultrasoft projectors were used for the 2s and 2p nonlocal components. An energy cut-off of 700 eV is used for the plane wave basis set expansion. Prior $^{19}$F chemical shielding calculations, symmetry-constrained molecular geometry optimizations are performed using PBE[61] functional (Table 1). The molecular state is simulated in CASTEP using a box large enough (1000 Å$^3$) to avoid interactions between molecular images.

For the calculation of $^{19}$F σ$_{iso}$ on crystalline systems (CASTEP code) twelve binary compounds are considered. Two structural data sets are used, the experimental structures reported in the literature (named IS in the following for initial structures) and the structures obtained after PBE-DFT atomic position optimization (APO structures) when allowed by symmetry which is only the case for MgF$_2$,[78] YF$_3$[79] and LaF$_3$.[80] Effectively, the alkali fluorides (LiF,[81] NaF,[82] KF,[83] RbF[84] and CsF[85]) adopt the NaCl structure type, three of the four studied alkaline earth basic fluorides (CaF$_2$,[86] SrF$_2$[87] and BaF$_2$[88]) adopt the fluorite structure type and ScF$_3$ adopts a ReO$_3$ type structure.[89,90] For these nine compounds, the atomic coordinates are therefore constrained by the local symmetry. To obtain converged $^{19}$F σ$_{iso}$ values, a plane wave basis set energy cut-off of 700 eV is necessary and a Monkhorst-Pack grid density approximately equal to 0.04 Å$^{-1}$ (corresponding to a *k*-point mesh of 8 × 8 × 8 for all structures except for YF$_3$ (4 × 4 × 6) and LaF$_3$ (4 × 4 × 4)) is enough. For the electronic loops, the PBE functional[61] is used for the exchange-correlation kernel. Total energies are converged up to changes smaller than 2 × 10$^{-5}$ eV. APO are obtained by minimizing the residual forces on the atom up to |F|$_{max}$ below 20 meV.Å$^{-1}$, keeping symmetry constraints and fixing the cell parameters to the experimentally determined values.

As previously proposed by Profeta *et al.*[60] for the Ca$^{2+}$ ion (3d$^0$), the local potentials of Sc$^{3+}$ (3d$^0$) and La$^{3+}$ (4f$^0$) USPP are also artificially shifted higher in energy compared to the default definition proposed by the Materials Studio package. This overcomes the deficiency of the PBE functional which generates too much covalent interaction between those empty states and the anionic *p* states. To show the limit of the PBE-DFT functional to describe these cations, calculation of the density of states (DOS) using hybrid PBE0 functional[91] are also performed for CaF$_2$. Norm-conserving pseudopotentials (NCPP) with a higher energy cut-off value (1088 eV) have to be used, USPP being not yet supported with hybrid functional.

EFG are calculated for $^{25}$Mg in MgF$_2$ and $^{139}$La in LaF$_3$ using the PAW[66] method implemented in CASTEP and the linearized augmented plane wave LAPW[69] method implemented in the WIEN2K package. The same PBE functional is used to compare calculated values of the EFG. The atomic sphere radii (R$_{MT}$) were set to 1.85 a.u. for Mg and F in MgF$_2$ and to 2.41 and 2.13 a.u. for La and F, respectively in LaF$_3$. Core states are 1s for Mg and F and from 1s to 4d for the La. The plane wave cut-off is defined by R$_{MT}$K$_{MAX}$ = 8. We use the same Monkhorst-Pack scheme as for CASTEP (8 × 8 × 8 for MgF$_2$ and 4 × 4 × 4 for LaF$_3$). Both sets of structures are used, the IS reported in the literature and the APO structures obtained using the CASTEP package.

**Conventions**

In this study, the calculated σ$_{iso}$ value is defined as:

$$\sigma_{iso} = (\sigma_{xx} + \sigma_{yy} + \sigma_{zz})/3,$$

σ$_{ii}$ being the principal components of the shielding tensor defined in the sequence | σ$_{zz}$ - σ$_{iso}$ | ≥ | σ$_{xx}$ - σ$_{iso}$ | ≥ | σ$_{yy}$ - σ$_{iso}$ |.

The isotropic chemical shift is defined as:

$$\delta_{iso} = - [\sigma_{iso} - \sigma_{ref}]$$

The quadrupolar coupling constant (C$_Q$) and the asymmetry parameter (η$_Q$) of the EFG tensor are defined as:

$$C_Q = (eQV_{zz}) / h,$$

$$\eta_Q = (V_{xx} - V_{yy}) / V_{zz}$$

V$_{ii}$ being the principal components of the EFG tensor defined in the sequence |V$_{zz}$| ≥ |V$_{yy}$| ≥ |V$_{xx}$|.

The quadrupolar moments (Q) of $^{25}$Mg and $^{139}$La are taken from ref [92].

## Results and discussion

### NMR shielding calculation on molecular systems: USPP validation

To validate the fluorine USPP used for the $^{19}$F σ$_{iso}$ GIPAW calculations, the GIPAW results are faced to AE calculations for a molecular benchmark. To ensure the computation accuracy of AE methods, we first compare the $^{19}$F σ$_{iso}$ calculated with GIAO and IGAIM methods and then $^{19}$F σ$_{iso}$ GIAO calculation results are compared to the values issued from the GIPAW method (Table 2).

The molecules used in our benchmark allows to validate our USPP on a large range of NMR shielding from highly shielded $^{19}$F atom (highly ionic C-F interaction on the CH$_3$F molecule) to much unshielded $^{19}$F atom (covalent F-F interaction on the F$_2$ molecule). Between both GIAO and IGAIM AE methods, we notice a convergence of the $^{19}$F σ$_{iso}$ values when increasing the accuracy of the basis sets from double zeta (aug-cc-pCVDZ) to quintuple zeta (aug-cc-pCV5Z), the differences between the two methods becoming negligible for the very large aug-cc-pCV5Z basis set. This ensures the validity of these AE references to test the fluorine USPP.

Table 2 $^{19}$F $\sigma_{iso}$ values (ppm) using different AE basis sets, with increasing accuracy, within the GIAO and IGAIM (in italic) methods. The last column reports the results obtained using USPP within the GIPAW method

| Molecule | All-electron | | | | Pseudopotential |
|---|---|---|---|---|---|
| | aug-cc-pCVDZ | aug-cc-pCVTZ | aug-cc-pCVQZ | aug-cc-pCV5Z | USPP |
| CH$_3$F | 453.2 | 452.6 | 452.1 | 451.9 | 452.1 |
| | *446.0* | *452.0* | *452.0* | *451.9* | |
| HF | 405.2 | 401.2 | 400.5 | 400.1 | 398.8 |
| | *401.6* | *401.3* | *400.5* | *400.1* | |
| C$_6$F$_6$ | 322.5 | 319.8 | 317.5 | 316.9 | 310.6 |
| | *320.0* | *319.0* | *317.6* | *316.8* | |
| CH$_2$F$_2$ | 310.7 | 304.4 | 302.4 | 301.6 | 298.7 |
| | *304.2* | *303.8* | *302.3* | *301.6* | |
| CF$_4$ | 225.0 | 216.8 | 213.9 | 212.6 | 207.0 |
| | *221.1* | *216.5* | *213.9* | *212.6* | |
| CFCl$_3$ | 138.2 | 125.8 | 121.9 | 120.3 | 113.2 |
| | *128.2* | *125.0* | *121.9* | *120.3* | |
| NF$_3$ | -36.7 | -50.1 | -56.2 | -58.8 | -73.5 |
| | *-40.4* | *-50.3* | *-56.2* | *-58.7* | |
| F$_2$ | -257.1 | -273.2 | -282.5 | -286.2 | -296.3 |
| | *-258.3* | *-273.6* | *-282.8* | *-286.3* | |

As reported by Ceresoli *et al.*,[55] the largest differences (from 10 to 15 ppm) between the AE methods and the USPP GIPAW calculations are observed for molecules having a highly negative $^{19}$F $\sigma_{iso}$ value. The weak shielding due to the covalent N-F or F-F interactions results in the contraction of the core orbitals. In this case, the number of Gaussian functions needed to correctly describe the atomic behaviour is a crucial parameter. This can be clearly seen in the evolution of the $^{19}$F $\sigma_{iso}$ values which still decrease by about 3.5 ppm for F$_2$ but only 0.2 ppm for CH$_3$F when increasing the basis set from quadruple zeta (aug-cc-pCVQZ) to quintuple zeta (aug-cc-pCV5Z). This would suggest that the calculated $\sigma_{iso}$ value is not yet completely converged with respect to the basis set size for the F$_2$ molecule. Such negative values for the shielding being not observed for crystalline systems, one can consider that the description of the core by GIPAW USPP is as good as the description by very large AE basis set. Probably GIPAW USPP calculations could be improved for much unshielded fluorine atoms by allowing core states relaxation (for the OTF_USPP generator) during the self consistent electronic procedure but this was beyond the scope of the present study.

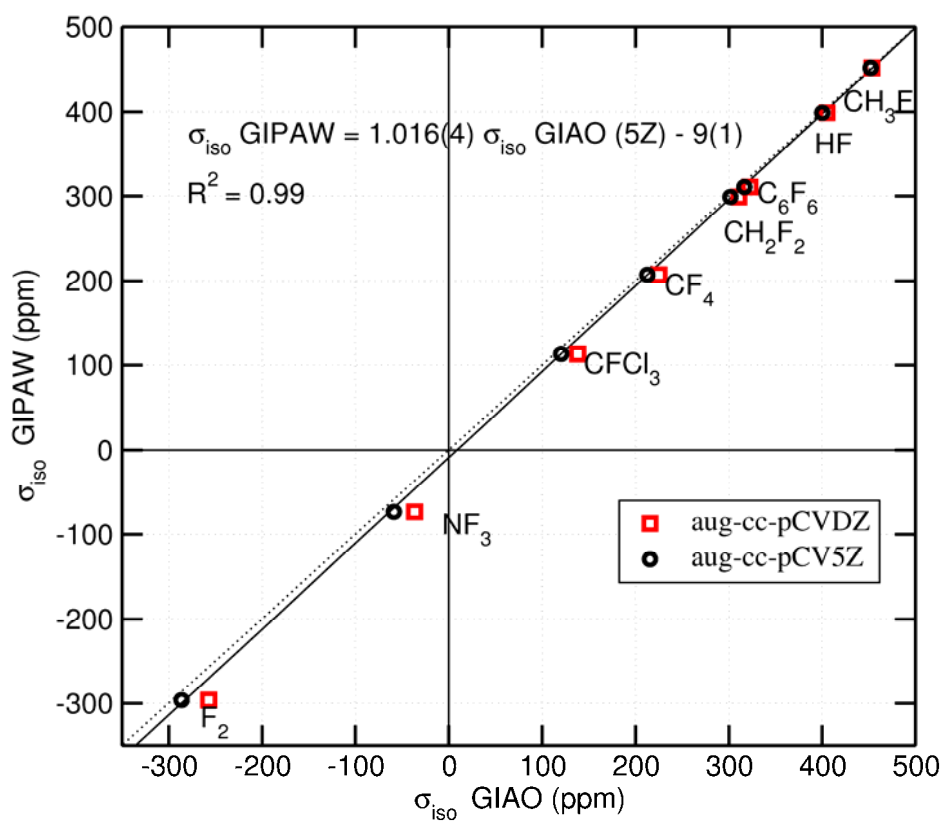

**Fig. 1** Calculated $^{19}$F $\sigma_{iso}$ values using USPP GIPAW method versus AE aug-cc-pCVDZ and aug-cc-pCV5Z basis sets with the GIAO method, using the PBE functional and the same molecular geometries (see Table 1 for details). The solid line represents the calculated linear regression corresponding to the equation reported on the graph for the aug-cc-pCV5Z basis set. The dotted line represents the ideal expected correlation $\sigma_{iso}$ GIPAW = $\sigma_{iso}$ GIAO.

Figure 1 shows the correlation between the $\sigma_{iso}$ values obtained with the all-electron GIAO method and the GIPAW method using the USPP of the Material Studio package. The remarkable agreement proves the correctness of the USPP fluorine atoms for calculating $^{19}$F $\sigma_{iso}$ when using PBE functional. They will be used in the following for crystalline systems.

**NMR shielding calculations on crystalline systems**

To perform a reliable comparison between experimental and DFT-GIPAW calculated $^{19}$F isotropic chemical shifts for alkali, alkaline earth and rare earth of column 3 basic fluorides, the consideration of accurate experimental values referenced relative to the same standard is a crucial point. Unfortunately, there are some discrepancies between the $^{19}$F $\delta_{iso}$ values previously reported for these compounds.[1,4-6,30,64,93-102] In addition, these values are referenced relative to different fluorine standard (CFCl$_3$, C$_6$F$_6$) and thus need to be expressed with respect to CFCl$_3$, the primary fluorine standard. Such conversion procedure was used in previous works[26,56] (some experimental values seem erroneously converted in the paper of Zheng et al.[56]. In order to obtain reliable data for comparison with calculations, we have therefore measured again the $^{19}$F $\delta_{iso}$ values for these fluorides with respect to CFCl$_3$ (measured $^{19}$F $\delta_{iso}$ are given in Table 3 and the corresponding experimental $^{19}$F NMR spectra are presented as ESI). For the compounds involving a single fluorine crystallographic site, $^{19}$F MAS NMR spectra were recorded at 7.0 T with MAS spinning frequency ranging from 15 to 30 kHz. For compounds containing several distinct F sites (YF$_3$ and LaF$_3$), a higher magnetic field of 17.6 T and fast MAS spinning frequency (up to 65 kHz) were employed to obtain very high resolution $^{19}$F MAS NMR spectra. The assignment of the NMR lines are unambiguous for the twelve studied compounds since they have only one fluorine site[78,81-90] or several fluorine sites with different multiplicities (2 for YF$_3$[79] and 3 for LaF$_3$[80]. Some difficulties were encountered with the determination of the $^{19}$F $\delta_{iso}$ value of ScF$_3$ due to local disorder related to its negative thermal expansion.[103] Experimental results on this compound are discussed in ESI.

A second point that must be considered with attention is the conversion of calculated $^{19}$F $\sigma_{iso}$ values into $^{19}$F $\delta_{iso}$ values. In the work of Zheng et al.[56], the calculated $^{19}$F $\sigma_{iso}$ values were converted into $^{19}$F $\delta_{iso}$ values with respect to C$_6$F$_6$ from the calculated $\sigma_{iso}$ value of the C$_6$F$_6$ reference molecule, as previously done by Yates et al.[104]. Then, the experimentally measured $^{19}$F $\delta_{iso}$ value of C$_6$F$_6$ relative to CFCl$_3$ is used to deduce "calculated" $^{19}$F $\delta_{iso}$ values relative to CFCl$_3$. Unfortunately, the experimental $^{19}$F $\delta_{iso}$ value of C$_6$F$_6$ used in references [56] and [104] are different and one can find in the literature several different $^{19}$F $\delta_{iso}$ values for C$_6$F$_6$ (relative to CFCl$_3$). To avoid this referencing problem and possible errors coming from the calculation of the $^{19}$F $\sigma_{iso}$ value of the isolated molecule chosen as reference, we have directly deduced the "calculated" $^{19}$F $\delta_{iso}$ values from the linear regression between calculated $^{19}$F $\sigma_{iso}$ values and experimental $^{19}$F $\delta_{iso}$ values referenced to CFCl$_3$.[58,59]

The $^{19}$F DFT-GIPAW $\sigma_{iso}$ values for alkali, alkaline earth and rare earth of column 3 basic fluorides calculated using the USSP and computational parameters presented in previous section and the corresponding measured $^{19}$F $\delta_{iso}$ are given in Table 3.

**Table 3** Experimental $^{19}$F $\delta_{iso}$ values, $^{19}$F $\sigma_{iso}$ values calculated using USPP within the GIPAW method for IS and APO structures, and calculated $\delta_{iso}$ values deduced from the linear regression obtained for YF$_3$, alkali and alkaline earth compounds without CaF$_2$ ($\delta_{iso}$/CFCl$_3$ = -0.80(3) $\sigma_{iso}$ + 89(9))

| Compounds | $\sigma_{iso}$ calc/ppm | | $\delta_{iso}$ calc/ppm | | $\delta_{iso}$ exp/ppm |
|---|---|---|---|---|---|
| | IS | APO | IS | APO | |
| LiF | 369.3 | - | -206 | - | -204.3(3) |
| NaF | 395.8 | - | -228 | - | -224.2(2) |
| KF | 268.1 | - | -125 | - | -133.3(2) |
| RbF | 221.3 | - | -88 | - | -90.9(2) |
| CsF | 136.3 | - | -20 | - | -11.2(2) |
| MgF$_2$ | 362.7 | 362.7 | -201 | -201 | -197.3(4) |
| CaF$_2$ | 220.0 / 246.2[a] | - | -87 / -108[a] | - | -108.0(2) |
| SrF$_2$ | 215.3 | - | -83 | - | -87.5(2) |
| BaF$_2$ | 151.9 | - | -33 | - | -14.3(2) |
| ScF$_3$ | 97.2 / 156.0[b] | - | 11 / -36[b] | - | -36(1) |
| YF$_3$ (F1) | 180.1 | 181.3 | -55 | -56 | -68.1(2) |
| YF$_3$ (F2) | 170.8 | 170.0 | -48 | -47 | -56.9(2) |
| LaF$_3$ (F1) | 93.7 / 133.6[c] | 91.8 / 132.1[c] | 14 / -18[c] | 15 / -17[c] | -23.6(2) |
| LaF$_3$ (F2) | 39.1 / 82.6[c] | 38.7 / 82.3[c] | 58 / 23[c] | 58 / 23[c] | 25.3(2) |
| LaF$_3$ (F3) | 47.2 / 89.3[c] | 52.5 / 94.2[c] | 51 / 18[c] | 47 / 13[c] | 16.9(2) |

[a] a shift of 1.81 eV was applied on the 3d orbitals
[b] a shift of 1.96 eV was applied on the 3d orbitals
[c] a shift of 4.55 eV was applied on the 4f orbitals

The linear correlation between experimental $\delta_{iso}$ and calculated $\sigma_{iso}$, from APO structures (see ESI) when allowed by symmetry, is shown Figure 2. Except for the F3 site in LaF$_3$, the $\sigma_{iso}$ values calculated from IS and APO structures are very similar. This is in agreement with slight optimization effects on F-Mg, F-Y or F-La distances (see ESI) and tends to show that these three structures were precisely determined. The slope of the linear regression (-0.70) is far below the theoretically expected value of minus one. However, same kind of deviations have been noted previously for other halogens (Cl, Br and I)[105-108] and other nuclei such as $^{29}$Si,[109,110] $^{31}$P,[111,112] $^{43}$Ca,[113] or $^{93}$Nb[114] and therefore does not seem to be a specific problem associated with fluorine NMR parameters. Similar trends were also reported by Zheng et al.[56] and by Griffin et al.,[26] with slopes equal to -0.86 and -0.68, respectively. This deviation to the theoretically expected slope of minus one already reported for PBE-DFT calculations implies establishing an empirical calibration curve to predict calculated isotropic chemical shift values. Another striking point is that the slope obtained here, which is relatively close to that obtained by Griffin et al.,[26] differs significantly from the slope reported by Zheng et al.[56] This difference arises mainly from the consideration of the calculated $\sigma_{iso}$ values for LaF$_3$, ScF$_3$ and CaF$_2$, two of these compounds containing cations ($Ca^{2+}$ and $La^{3+}$) already known to be inaccurately described with PBE-DFT.[60,115] In the work of Griffin et al., two of these three compounds ($CaF_2$ and $LaF_3$) were considered while only $CaF_2$ was studied by Zheng et al. leading to a larger absolute slope.

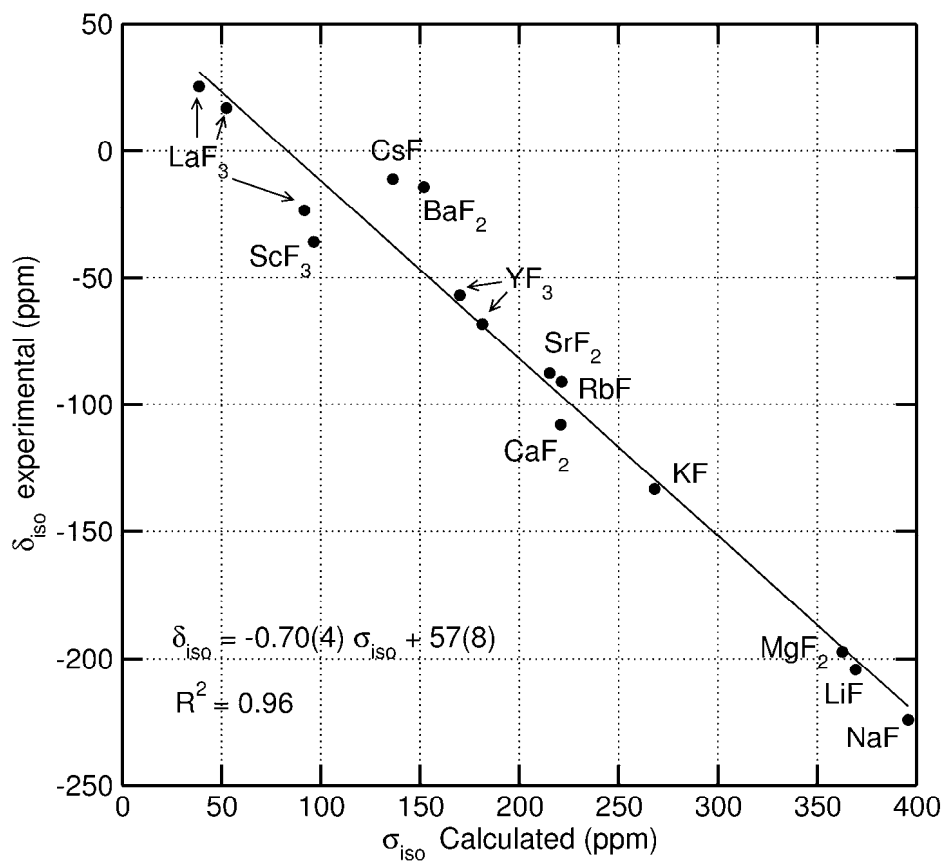

**Fig. 2** Calculated $^{19}F$ $\sigma_{iso}$ values using PBE functional for APO structures when allowed by symmetry versus experimentally measured $^{19}F$ $\delta_{iso}$ values. The solid (black) line represents the calculated linear regression..

The deficiency of PBE-DFT in calculating the NMR shielding of anions neighboured by $Ca^{2+}$ cations has already been reported by Profeta et al.[60]. They have shown that the PBE functional leads to an inaccurate calculation of $^{17}O$ $\sigma_{iso}$ in CaO due to an overestimation of the Ca-O bond covalence. More precisely, too much interaction is found by PBE between the Ca(3d) and O(2p) states. To overcome this PBE-DFT deficiency, the energy level of the 3d Ca orbitals was shifted to higher energy without changing the position of the s and p states in the Ca pseudopotential.[60] This method has been afterwards successfully applied on $^{43}Ca$ NMR parameters calculations.[116] Following these works, we have applied an empirical shift on the Ca(3d) orbitals for building the Ca USPP. To determine the optimal 3d-shift a reference correlation between $\delta_{iso}$ and $\sigma_{iso}$ is needed and we have thus established a new correlation presented in figure 3 by discarding $CaF_2$, $ScF_3$ and $LaF_3$.

It leads to:

$\delta_{iso}/CFCl_3 = -0.80(3) \sigma_{iso} + 89(9)$ (equation 1)

From this new linear regression (equation 1), the ideal $^{19}F$ $\sigma_{iso}$ value for the fluorine in $CaF_2$ can be established (Figure 3(a)) and further used to adjust the 3d-shift for the Ca(3d) orbitals. An optimal shift of 1.81 eV is obtained (see Figure 3(b)). This value is significantly smaller than for CaO (3.2 eV).[60] This difference can be explained as (i) we used USPP whereas Profeta et al. used NCPP[60] and (ii) the degrees of covalency of the Ca-O and Ca-F bonds are different. To ascertain this empirical procedure, the density of states (DOS) obtained using PBE functional for the two different definitions of the Ca USPP is compared with the DOS obtained using hybrid functional PBE0, which is expected to give better description of the covalency in the system (Figure 4).

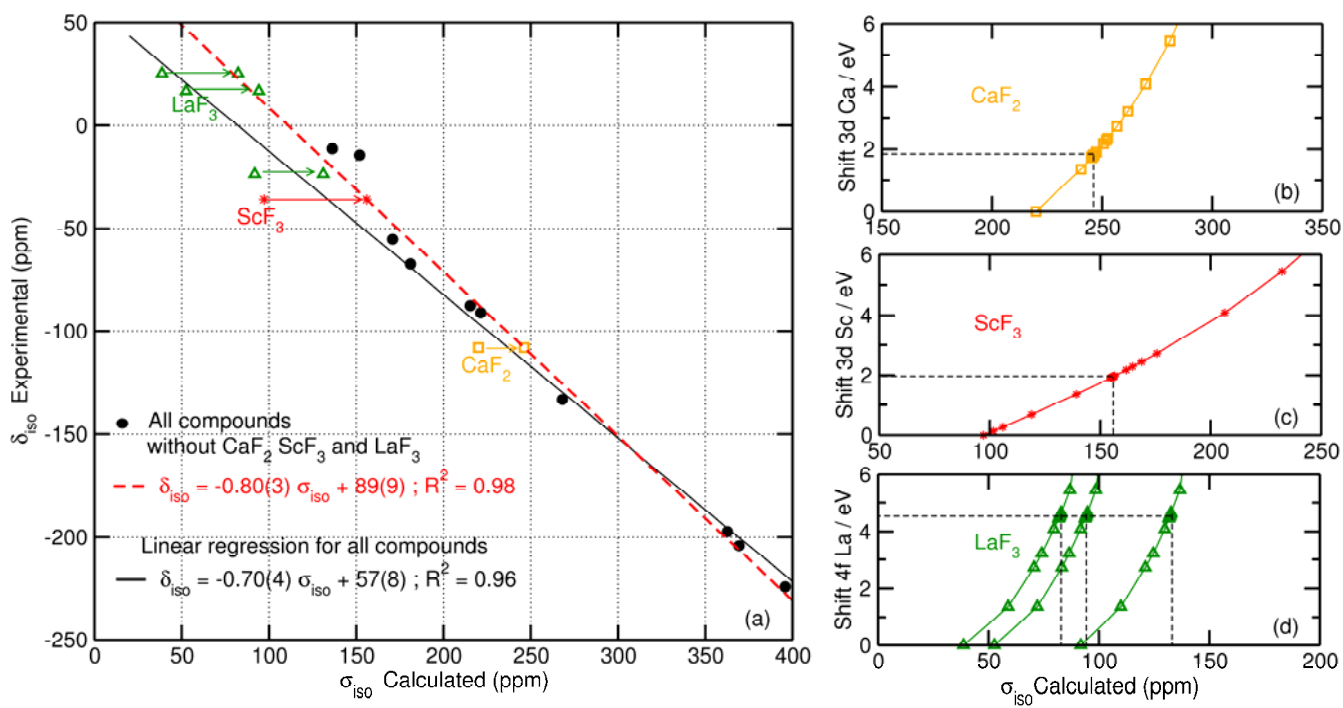

**Fig. 3** (a) Experimental $^{19}$F $\delta_{iso}$ (/CFCl$_3$) values versus calculated $\sigma_{iso}$ values. The solid (black) line represents the linear regression when considering all compounds and the dashed (red) line represents the linear regression when considering YF$_3$ and alkali and alkaline earth compounds without CaF$_2$. The arrows represent the change in $\sigma_{iso}$ when applying a shift on the 3d orbitals of Ca and Sc and on the 4f orbitals of La. The panels on the right side report the $\sigma_{iso}$ evolution with the applied shifts on the 3d orbitals of Ca (b) and Sc (c) and on the 4f orbitals of La (d).

The effect of the shift applied on the 3d orbitals is clearly observed in the conduction band: the energy of the band having mostly a Ca(3d) character is increased and becomes closer to the one obtained using hybrid functional. The band gap stays unchanged (mainly imposed by the position of Ca(4s) states in the conduction band) and is calculated to 6.3 eV using PBE functional. As expected, the use of hybrid functional gives a higher band gap value (8.4 eV) which is closer to the experimental one (11.8 eV).[117] Since the calculated $^{19}$F $\sigma_{iso}$ of ScF$_3$ also deviates significantly from the linear regression established for YF$_3$, alkali and alkaline earth compounds excluding CaF$_2$ (as evidenced from Figure 3(a)) and because DOS calculation (not shown) shows that the bottom of the conduction band has a strong 3d character, it appears that a similar correction is also needed to properly describe the 3d orbitals of the Sc atom when using PBE functional. As done for Ca, we thus adjusted the corresponding 3d-shift for the Sc USPP such that the calculated $^{19}$F $\sigma_{iso}$ corresponds to the value determined from the experimental $\delta_{iso}$ using equation (1). Figure 3(c) shows that the effect of the applied 3d-shift on the calculated $\sigma_{iso}$ value is more pronounced for ScF$_3$ than for CaF$_2$. It should be noticed that the values of the 3d-shift required for the Sc$^{3+}$ ion in ScF$_3$ (1.96 eV) and for the Ca$^{2+}$ ion in CaF$_2$ (1.81 eV) are very close. This observation gives some confidence about the relevance of this empirical procedure to overcome the deficiency of PBE functional in describing cations with localized 3d empty states.

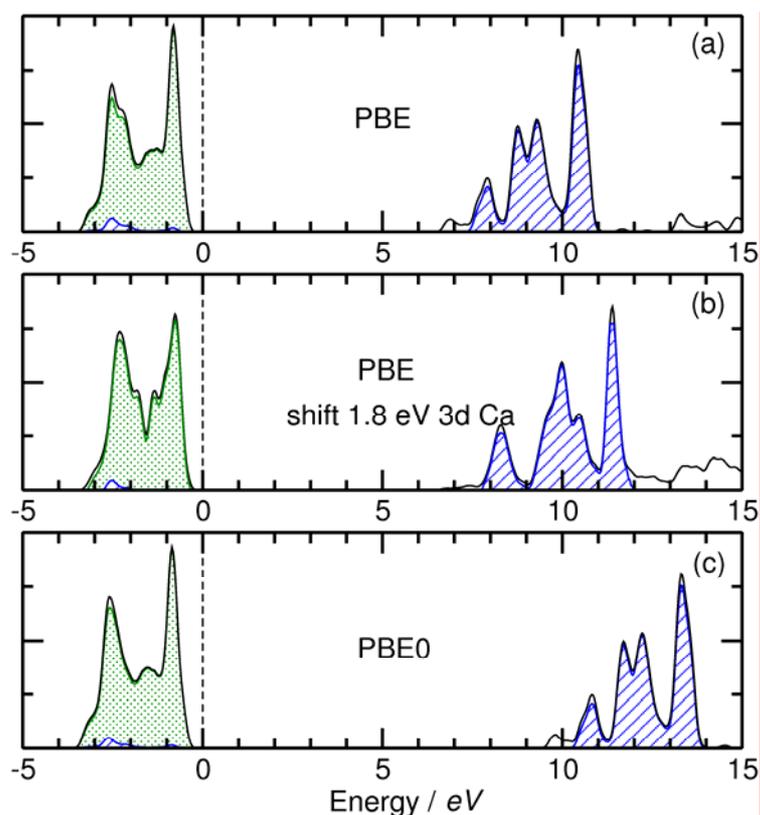

**Fig. 4** DOS for $CaF_2$ using (a) PBE functional, (b) PBE functional with a 3d-shift of 1.81 eV for Ca and (c) PBE0 hybrid functional. The blue area represents the partial density of states projected on the 3d orbitals of the Ca element and the green area the partial density of states projected on the 2p orbitals of the F element.

It is also known that standard GGA/DFT is not well suitable to elements with localized 4f empty states. For example, a recent theoretical investigation has shown that it is necessary to add an on-site Hubbard correction ($U_{eff}$ = 10.3 eV) on the 4f(La) orbitals to properly describe their localizations and then their energy positions, allowing to properly simulate the XPS/BIS and reflectance experimental spectra of $LaF_3$.[115] In our case, the very large deviation observed for $LaF_3$ (Figure 3(a)) shows that the $La^{3+}$ ion ($4f^0$) has a similar symptomatic behaviour as $Ca^{2+}$ and $Sc^{3+}$ ions. Therefore, we have applied the Profeta *et al.*[60] procedure to shift 4f orbitals. $LaF_3$ having three fluorine sites, the 4f-shift was determined by simultaneously minimizing for the three sites the differences between the experimental $^{19}F$ $\delta_{iso}$ and the $\delta_{iso}$ values deduced from the calculated $^{19}F$ $\sigma_{iso}$ using equation (1). The optimum value obtained following this protocol (4.55 eV, Figure 3(d)), is much higher than the one determined for the 3d orbitals of $Ca^{2+}$ and $Sc^{3+}$.

The validity of equation (1), used to determine the 3d- and 4d-shifts required to calculate the $^{19}F$ $\sigma_{iso}$ of compounds for which the lowest energy states of the conduction bands have strong 3d or 4d characters (*i.e.* $CaF_2$, $ScF_3$ and $LaF_3$), is illustrated in figure 5. In this plot, we have reported the $^{19}F$ $\delta_{iso}$ values previously measured for several other inorganic fluorides together with the corresponding $^{19}F$ $\sigma_{iso}$ values calculated by Zheng *et al.*[56] and by Griffin *et al.*[26] using the PBE-DFT GIPAW method with the same fluorine USPP (see ESI). Great care was taken to consider only compounds for which $^{19}F$ $\delta_{iso}$ values were determined from high-resolution spectra (*i.e.* recorded at relatively high magnetic fields using MAS spinning frequency larger than 10 kHz) and for which an unambiguous assignment of the resonances is provided. These compounds include $ZnF_2$,[30] $CdF_2$,[30] $\alpha$-$PbF_2$ (2 distinct F sites),[118] $HgF_2$,[101] $\alpha$-$AlF_3$,[119] $GaF_3$,[30] $InF_3$,[30] $BaLiF_3$,[30] $Na_5Al_3F_{14}$[20] (3 distinct F sites), $\beta$-$BaAlF_5$[17] (10 distinct F sites) and $Ba_3Al_2F_{12}$[17,29] (8 distinct F sites). It should be mentioned that the GIPAW calculations of the $^{19}F$ isotropic shielding by Griffin *et al.*[26] and Zheng *et al.*[56] were carried out using slightly different computation parameters than those used in this work. Griffin *et al.*[26] have used a cut-off energy of 680 eV and a k-spacing of 0.04 Å$^{-1}$, and a full geometry optimization (variation of both the lattice parameters and internal atomic coordinates) was performed prior the $\sigma_{iso}$ calculations. Zheng et al.[56] have employed much smaller cut-off energies of 300 eV for the optimizations of atomic positions and of 550 eV for the GIPAW calculations, which does not allow obtaining fully converged $^{19}$F $\sigma_{iso}$ values (see experimental section). Taking these aspects into account, it is clearly observed in Figure 5 that equation (1) ($\delta_{iso}$/CFCl$_3$ = -0.80(3) $\sigma_{iso}$ + 89(9)) fits perfectly with these results obtained for other inorganic fluorides, the linear regression obtained considering these 11 compounds in addition to the 12 studied fluorides being $\delta_{iso}$/CFCl$_3$ = -0.79(1) $\sigma_{iso}$ + 90(3)).

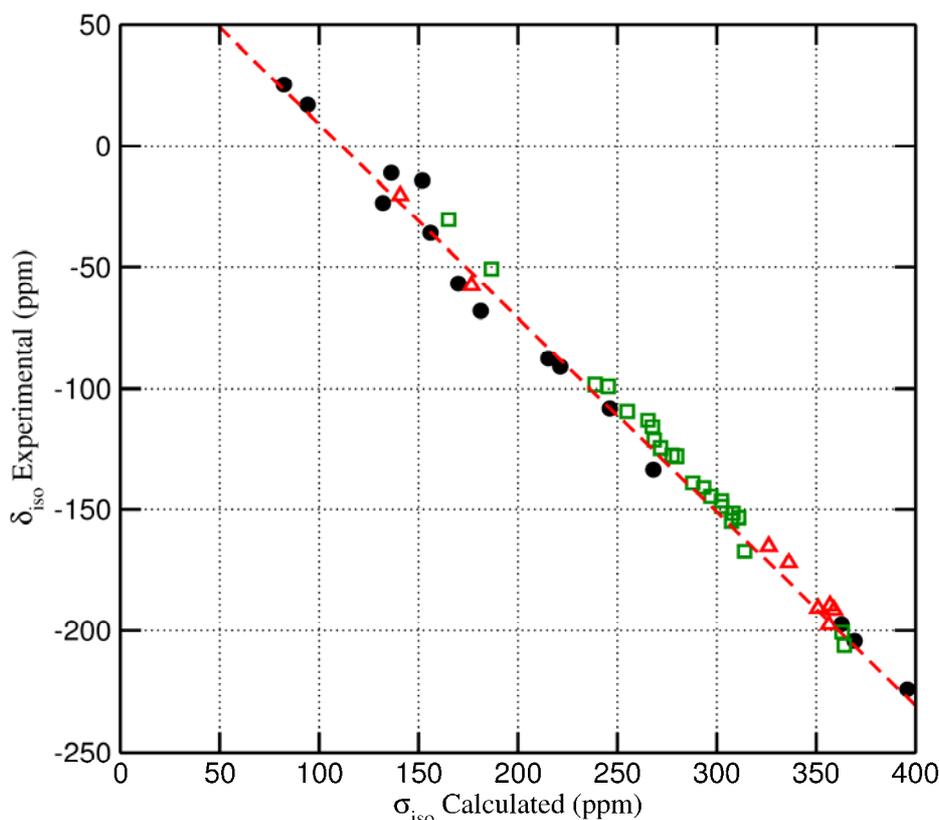

**Fig. 5** Calculated $^{19}$F $\sigma_{iso}$ values using PBE functional versus experimentally measured $^{19}$F $\delta_{iso}$ values. The dark circles represent the values reported in Table 3. The red triangles represent the values calculated by Griffin et al.[26] for CdF$_2$, HgF$_2$, α-PbF$_2$, α-AlF$_3$ and Na$_5$Al$_3$F$_{14}$. The green squares represent the values calculated by Zheng et al.[56] for ZnF$_2$, GaF$_3$, InF$_3$, BaLiF$_3$, β-BaAlF$_5$ and Ba$_3$Al$_2$F$_{12}$. The values presented on this figure are reported as ESI. The dashed (red) line corresponds to equation (1) ($\delta_{iso}$/CFCl$_3$ = -0.80(3) $\sigma_{iso}$ + 89(9)).

By applying equation (1) to the calculated $^{19}$F $\sigma_{iso}$ values plotted in Figure 5 (except those of CaF$_2$, ScF$_3$ and LaF$_3$ for which this equation was used to adjust the 3d- and 4f-shifts), a standard deviation between experimental and "calculated" $^{19}$F $\delta_{iso}$ of 7 ppm is obtained. This indicates that this equation can be used to predict the $^{19}$F NMR spectra of crystalline compounds from PBE-DFT GIPAW calculation with a quite good accuracy. More importantly, it should be pointed out that for all the compounds having multiple fluorine crystallographic sites (2 sites in YF$_3$, 3 sites in LaF$_3$, 2 sites in α-PbF$_2$, 10 sites in β-BaAlF$_5$ and 8 sites in Ba$_3$Al$_2$F$_{12}$), the relative positions of the calculated $\sigma_{iso}$ values (and the corresponding "calculated" $\delta_{iso}$ values) is similar to the relative positions of the experimental $\delta_{iso}$ values, showing that such calculations allow an unambiguous assignment of $^{19}$F resonances for compounds having several fluorine sites with the same multiplicity.

### Electric field gradient calculations

In the second part of this work, we compare the calculated EFG tensor to the one experimentally determined using solid-state NMR. According to the symmetry of the cationic sites in the studied compounds, the EFG tensors of the quadrupolar nuclei occupying cationic sites are expected to be different from zero only for $^{25}$Mg in MgF$_2$ and $^{139}$La in LaF$_3$. The quadrupolar parameters $C_Q$ and $\eta_Q$ are directly related to the principal components of the EFG tensor which originates from the deformation of the electronic density around the nucleus. Consequently, $C_Q$ and $\eta_Q$ parameters are very sensitive to the site symmetry and/or site distortion and provide additional structural information. The quadrupolar parameters for $^{25}$Mg in MgF$_2$ which adopts the rutile structure type (P4$_2$/mnm space group),[78] are measured for the first time. These parameters and those of $^{139}$La in LaF$_3$ determined by Ooms et al.[65] and Lo et al.[64] are gathered in Table 4. The EFG tensor components calculated with both CASTEP and WIEN2K codes for the IS and APO structures of MgF$_2$ and LaF$_3$ are also reported in Table 4, together with the quadrupolar parameters of $^{139}$La in LaF$_3$ previously calculated by Ooms et al.[65] using LAPW method.[69]

Table 4 Experimental $C_Q$ and $\eta_Q$, calculated $V_{ii}$, $C_Q$ and $\eta_Q$ using CASTEP and WIEN2K for initial and APO structures. Since only the absolute value of $C_Q$ can be determined from NMR experiments on powdered samples, the sign of the experimental $C_Q$ is set to the sign of the calculated $C_Q$. The quadrupolar moment Q values are equal to 0.1994 x 10$^{28}$ m² and 0.2000 x 10$^{28}$ m² for $^{25}$Mg and $^{139}$La, respectively.[a]

|  |  | $V_{zz}$/10$^{21}$ V m$^{-2}$ | $V_{yy}$/10$^{21}$ V m$^{-2}$ | $V_{xx}$/10$^{21}$ V m$^{-2}$ | $C_Q$/MHz | $\eta_Q$ |
|---|---|---|---|---|---|---|
| MgF$_2$ |  |  |  |  |  |  |
| exp |  | 0.728(6) | -0.480(8) | -0.248(8) | 3.51(3) | 0.32(2) |
| IS | CASTEP | 0.631 | -0.434 | -0.196 | 3.04 | 0.38 |
|  | WIEN2K | 0.637 | -0.498 | -0.139 | 3.06 | 0.56 |
| APO | CASTEP | 0.655 | -0.364 | -0.291 | 3.16 | 0.11 |
|  | WIEN2K | 0.658 | -0.431 | -0.228 | 3.17 | 0.31 |
| LaF$_3$ |  |  |  |  |  |  |
| exp |  | -3.29(1) | 2.99(8) | 0.30(8) | -15.90(5)[b] | 0.82(5)[b] |
|  |  | -3.309 | 2.994 | 0.314 | -16.0[c] | 0.81[c] |
| IS | WIEN2K | -3.311 | 2.831 | 0.480 | -16.01[b] | 0.71[b] |
| IS | CASTEP | -3.147 | 2.755 | 0.393 | -15.22 | 0.75 |
|  | WIEN2K | -3.307 | 2.963 | 0.344 | -15.99 | 0.79 |
| APO | CASTEP | -3.722 | 3.432 | 0.290 | -18.00 | 0.84 |
|  | WIEN2K | -3.947 | 3.678 | 0.269 | -19.09 | 0.86 |

[a] from ref 92, [b] from ref 65, [c] from ref 64

As shown in Figure 6, the $^{25}$Mg MAS NMR spectra of MgF$_2$ recorded at two different magnetic fields (9.4 and 17.6 T) exhibit typical second order quadrupolar broadened line shapes. Good fits of the two $^{25}$Mg experimental spectra can be obtained taking into account only the second order quadrupolar interaction indicating that the effect of the $^{25}$Mg chemical shift anisotropy can be neglected even at 17.6 T. The $^{25}$Mg isotropic chemical shift determined from the fits of experimental spectra is -4±1 ppm. The calculated $\delta_{iso}$ deduced from the isotropic shielding $\sigma_{iso}$ calculated with CASTEP (564.6 ppm) using the equations reported by Pallister et al.[63] ($\delta_{iso}$ = -0.933 $\sigma_{iso}$ + 528.04) and Cahill et al.[62] ($\delta_{iso}$ = -1.049 $\sigma_{iso}$ + 565.23) are respectively equal to 1.4 ppm and -0.6 ppm which are both in fine agreement with the experimental value.

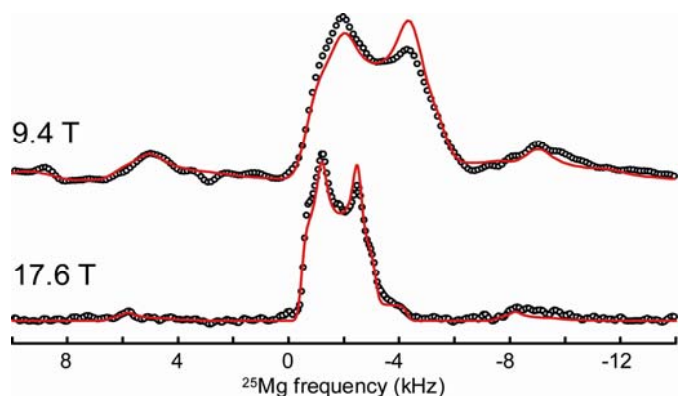

**Fig. 6** Experimental (black dots) $^{25}$Mg MAS (7 kHz) Hahn echo NMR spectra of MgF$_2$ recorded at 9.4 T (top) and 17.6 T (bottom) and their best fits (red lines).

As reported in Table 4, there are slight discrepancies between the measured $V_{ii}$ values and those calculated for the experimental structure of $MgF_2$ using the PAW USPP (CASTEP) or LAPW AE (WIEN2K) methods, the calculated $V_{zz}$ and $|V_{xx}|$ principal components of the EFG tensor being underestimated. As previously done for the $^{19}F$ chemical shielding, the principal components of the EFG tensors were also calculated for the APO structure. The PBE-DFT optimisation of the fluorine atomic position leads to slight modifications of the $MgF_6$ octahedron: the mean Mg-F distance remains the same (1.982 Å) but the radial distortion increases and the angular distortion decreases (see ESI). For the APO structure, a better agreement between experimental and calculated $V_{zz}$ values is obtained. Nevertheless, the $V_{xx}$ and $V_{yy}$ components calculated with the PAW USPP method are respectively larger and smaller than the experimental values leading to a discrepancy between the calculated and experimental asymmetry parameters ($\eta_Q$). In contrast, the $V_{ii}$ values (and thus $C_Q$ and $\eta_Q$ parameters) calculated using the LAPW AE method are in very good agreement with the experimental ones.

As shown in Figure 7(a) which depicts the orientation of the $^{25}Mg$ EFG tensor (see ESI for details), the $V_{ii}$ components are along the intersections of the three mirror planes of the Mg site (mmm symmetry) and, for the Mg atom located at (0,0,0), $V_{zz}$ and $V_{xx}$ lie in the (a,b) plane while $V_{yy}$ is along the c crystallographic axis. In the $MgF_2$ structure, the $MgF_6$ octahedron is characterized by low radial and high angular distortions (see ESI). In such a situation, the largest component of the EFG tensor ($V_{zz}$) is not expected to be oriented along M-F bonds.[120] Indeed, $V_{zz}$ and $V_{yy}$ are both oriented between two Mg-F bonds in the plane presenting the angular distortion while $V_{xx}$ is oriented along the shortest Mg-F bond perpendicular to this plane. It should also be noted that the sign of the calculated $V_{ii}$ components is in agreement with the angular distortion analysis model proposed by Body et al.,[120] i.e. a positive/above 90° (negative/below 90°) angular distortion leads to a charge depletion (concentration) in the $V_{ii}$ direction and then to a positive (negative) $V_{ii}$ value (Table 4). More detailed information about the origin of the EFG at the nucleus is traditionally obtained from charge density distribution visualized on electron density difference $\Delta\rho$ maps.[120-122] $\Delta\rho$ represents the difference between the crystalline electron density and the superposition of electron densities from the neutral atoms. The $\Delta\rho$ maps for the plane containing the $V_{xx}$ component and four fluorine atoms with Mg-F distances of 1.975 (x2) and 1.986 Å (x2) and for the plane containing $V_{zz}$ and $V_{yy}$ are shown in Figure 7(c) and 7(d), respectively. The $V_{xx}$ component being rather small the corresponding charge deformation is not easily evidenced. On the other hand, in the $V_{zz}/V_{yy}$ plane (Figure 7(d)), the expected depletion of charge in the $V_{zz}$ direction (i.e. between the two Mg-F bonds which form an angle equal to 98.8°) relative to the $V_{yy}$ direction (i.e. between the two Mg-F bonds which form an angle equal to 81.2°) is clearly observed. Isolines on the $\Delta\rho$ map close to the nucleus are effectively slightly compressed (elongated) along the $V_{zz}$ ($V_{yy}$) direction due to this depletion (increase) of the electronic density. As mentioned above, the $\Delta\rho$ maps are not well suited to establish a relationship between the EFG and the electronic density asphericity near the nucleus when the distortion is small. Another approach proposed by Schwarz et al.[123] consists in calculating the difference with respect to the ionic spherical density. However, this approach requires the construction of isolated ions (F$^-$, Mg$^{2+}$) which is not always straightforward. For simplicity we only consider the non-spherical contribution of the electronic density inside the Mg sphere (Figure 7(b)). One can observe that $V_{zz}$, which is positive, is oriented along the negative part of this non spherical density. We also note that the asymmetry of the positive part of the density which is less important along the $V_{xx}$ direction than along the $V_{yy}$ direction is in agreement with the lower absolute value of $V_{xx}$ compared to $V_{yy}$.

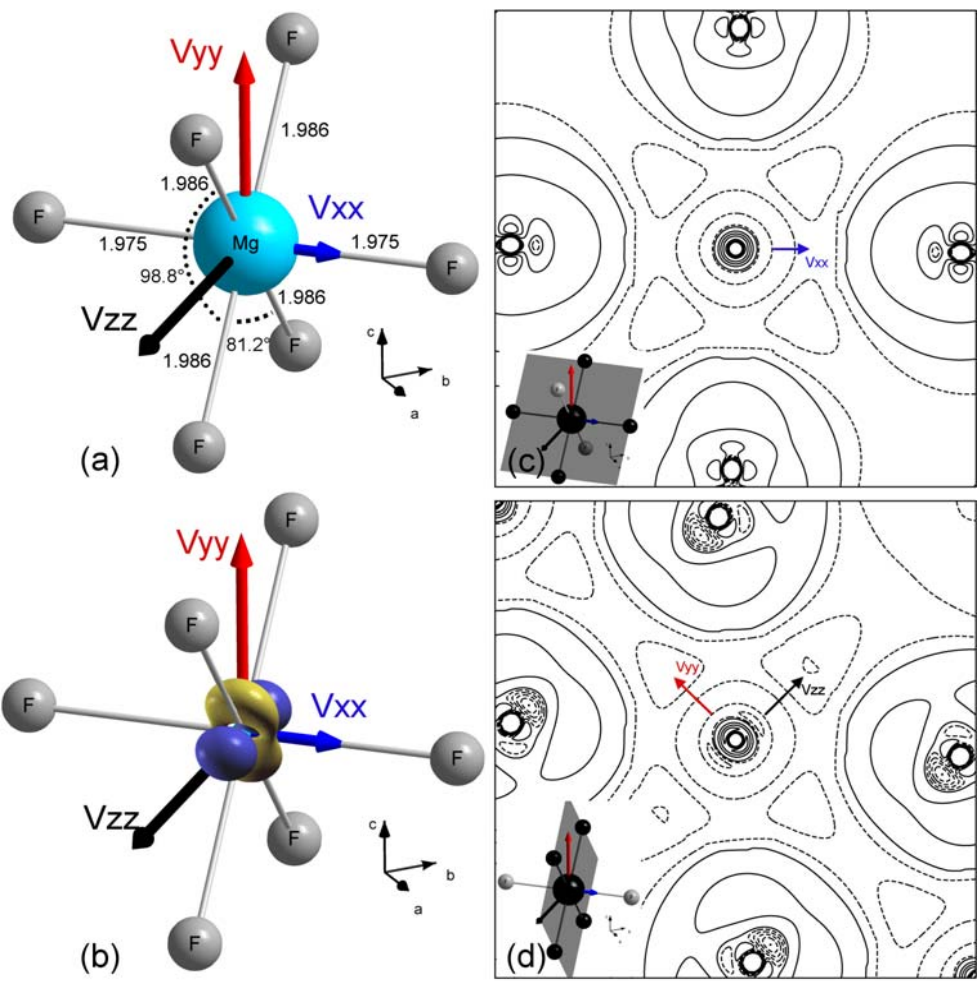

**Fig. 7** (a) Orientation of the $^{25}$Mg EFG tensor components, calculated with WIEN2K on the APO structure, represented on MgF$_6$ octahedron. Mg-F bond lengths (Å) and F-Mg-F bond angles (°) are indicated. The norms of the eigenvectors are proportional to the eigenvalues of the EFG tensor components (see Table 4). (b) Isosurface of the quadrupolar charge deformation considering only the |L| = 2 terms inside the Mg sphere. Light (yellow) colour is used for positive values and dark (blue) colour for negative values. For graphical convenience the volume has been increased by more than an order of magnitude. (c) Δρ map in the plane containing the V$_{xx}$ component and four Mg-F bonds. (d) Δρ map in the plane containing the V$_{zz}$ and V$_{yy}$ components. On these maps, solid and dashed lines represent respectively positive (from 0.002 to 0.065 e/a.u.$^3$ with a step of 0.016 e/a.u.$^3$) and negative (from -0.002 to -0.040 e/a.u.$^3$ with a step of 0.004 e/a.u.$^3$) values of the electronic density.

The trigonal structure of LaF$_3$ (*P*-3*c*1 space group) contains a single La crystallographic site (6f Wyckoff position).[80] The La coordination polyhedron is made of 9 fluorine atoms with La-F distances ranging from 2.417 to 2.636 Å and 2 additional fluorine atoms at a longer La-F distance of ~3 Å (see Figure 9 and ESI). The La site has a twofold symmetry axis which lies along the La-F3 bond, parallel to the crystallographic a-axis. As already reported,[65] the $^{139}$La quadrupolar parameters calculated for the experimental structure using the PAW USPP or LAPW AE methods are very close to the measured C$_Q$ and η$_Q$ values and the best agreement is obtained for the LAPW AE method (see Table 4). In contrast, some discrepancies between the V$_{ii}$ values calculated for the APO structure and the experimental ones are observed, the two computation methods leading to a notably overestimated V$_{zz}$ and V$_{yy}$ components. These overestimations of the values calculated after the geometry optimization step remains difficult to explain since the variation of the La environment, which is difficult to analyze for this coordination polyhedron, is small considering the La-F distances (see ESI). It should be pointed out that the initial structure of LaF$_3$ was determined with a very high accuracy (neutron diffraction on single crystal[80] and, in such a case, the weak variation of the structural parameters induced by the PBE-DFT geometry optimization leads to less accurate calculated EFG values.

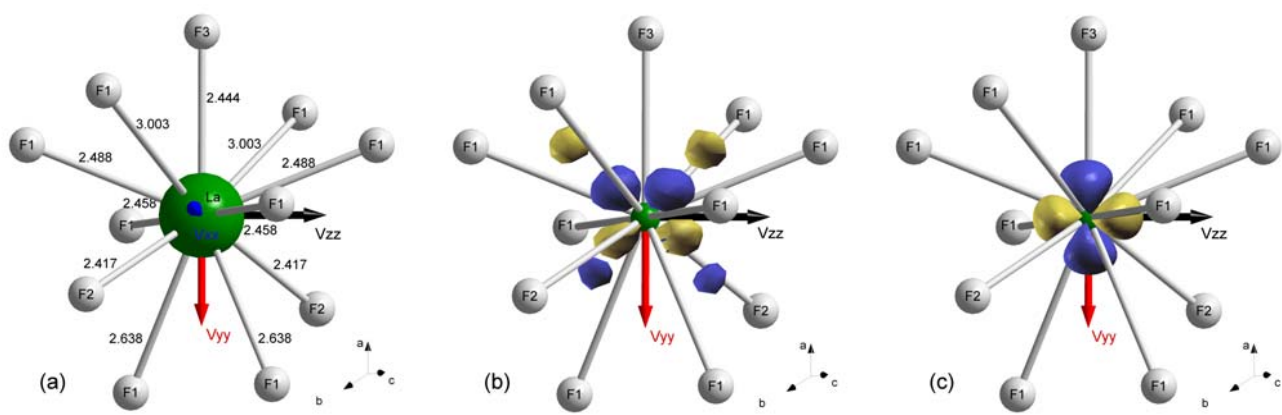

**Fig. 8** Orientation of the [139]La EFG tensor components, calculated with WIEN2K, on the initial structure, represented on LaF$_{11}$ polyhedron. The lengths of the vectors are proportional to the eigenvalues of the EFG tensor. (a) The bond lengths/Å are reported and the V$_{xx}$ component is multiplied by ten for graphical visualization (see Table 4). Representations of the electronic densities inside the La sphere: (b) reports the contributions to the density when the spherical terms have been removed and (c) reports the contributions to the density when only the |L| = 2 terms are considered. Light (yellow) colour is used for positive values and dark (blue) colour for negative values. For graphical convenience the volume has been increased by more than an order of magnitude.

As shown in Figure 8(a), V$_{yy}$ is oriented along the twofold symmetry axis (*i.e.* lies along the La-F3 bond). The complexity of the La environment prevents predicting the relative orientation of the [139]La EFG tensor components from simple coordination polyhedron geometry considerations,[120] and to go further in the analysis, electronic density differences were calculated. In a first step, the spherical part of the charge density is removed keeping all non-spherical terms in the LM expansion (Figure 8(b)). However, this does not allow finding a correlation between the electronic density and the orientation of the EFG eigenvalues. In a second step, only the |L| = 2 terms of the LM expansion is considered (Figure 8(c)) and a good correlation between the electronic density deformation and the orientation of the EFG eigenvectors is then found. A depletion of charge is observed along the twofold axis in agreement with the positive value of V$_{yy}$. Along this direction, the number of neighbouring fluorine atoms is rather small with longer La-F bonds (La-F1 bond lengths equal to 2.638 Å and 3.003 Å). In contrast, an accumulation of charge is observed along the V$_{zz}$ direction, in agreement with the negative value of V$_{zz}$. Along this direction, the number of neighbouring fluorine atoms is larger with shorter La-F bonds (La-F1 bond lengths equal to 2.458 Å and 2.488 Å and La-F2 bond lengths equal to 2.417 Å) leading to stronger La-F interactions. Finally, it should be noted that the depletion of charge in the V$_{yy}$ direction and the accumulation of charge in the V$_{zz}$ direction have similar amplitudes in agreement with similar absolute values of these eigenvalues ($\eta_Q$ value close to 1) and accordingly, only a tiny deformation of the electronic density is observed along the direction of V$_{xx}$ (Figure 8(c)).

## Conclusion

We have investigated the relationship between experimental [19]F $\delta_{iso}$ and calculated [19]F $\sigma_{iso}$ values from first-principles calculations using the GIPAW method and the PBE functional, for alkali, alkaline earth and rare earth of column 3 fluorides. On this basis, we show that the PBE functional is unable to reproduce the measured [19]F $\delta_{iso}$ value in CaF$_2$ as it overestimates the Ca-F covalence but this deficiency is corrected by applying a shift on the 3d orbitals. We also evidence that the same type of correction is required in the case of ScF$_3$ and LaF$_3$ for which the bottom of the conduction band has a strong 3d and 4f character, respectively, and we have determined the shifts of the 3d(Sc) orbitals and 4d(La) orbitals needed to accurately calculate the [19]F shielding tensors of these compounds using the PBE functional. Taking into account this deficiency of the PBE functional, we propose a correlation between the calculated [19]F $\sigma_{iso}$ values and the experimental [19]F $\delta_{iso}$ values that allow the prediction of [19]F NMR spectra with a relatively good accuracy. Nevertheless, our results highlight the need of to compute the NMR shielding using improved exchange-correlation functionals such as hybrid functionals. In this context, the converse approach recently developed by Thonhauser *et al.*[54] seems to be a promising solution. In this work, we also determined and calculated the quadrupolar parameters of $^{25}$Mg in MgF$_2$ and, from the analysis of charge distribution through electron density maps, it is shown that the orientation of the EFG components of $^{25}$Mg reflects the angular distortion of the MgF$_6$ octahedron. Finally, we have shown that the electronic density deformation determined by considering only the $|L| = 2$ terms of the LM expansion gives a reliable picture of the EFG tensors of $^{25}$Mg in MgF$_2$ and $^{139}$La in LaF$_3$.

## Acknowledgments

The authors are grateful for the financial support of the Région Pays de la Loire of the RMN3MPL project, especially M. Biswal (doctoral grant) and A. Sadoc (post-doctoral fellowship). Financial support from the TGIR RMN THC FR3050 is also gratefully acknowledged. The computational presented in this work have been carried out at the Centre Régional de Calcul Intensif des Pays de la Loire (CCIPL), financed by the French Research Ministry, the Région Pays de la Loire, and Nantes University. We thank CCIPL for CASTEP licenses financial support.

We also thank Cyrille Galven (Laboratoire des Oxydes et Fluorures), Alain Demourgues and Etienne Durand (ICMCB-Pessac) for their help in the study of ScF$_3$ (XRPD and fluorination experiments).

## Supporting Information Available

Experimental $^{19}$F MAS NMR spectra of alkali fluorides, alkaline earth fluorides, YF$_3$ and LaF$_3$. XRPD pattern and $^{19}$F MAS NMR spectrum of ScF$_3$. Atomic coordinates of IS and APO structures of MgF$_2$, YF$_3$ and LaF$_3$. Mg-F bond lengths and F-Mg-F bond angles determined from the IS and APO structures of MgF$_2$. Y-F bond lengths determined from the IS and APO structures of YF$_3$. La-F bond lengths determined from the IS and APO structures of LaF$_3$. Experimental $^{19}$F $\delta_{iso}$ and calculated $^{19}$F $\sigma_{iso}$ values presented in Figure 5. Eigenvectors of the calculated EFG tensors of $^{25}$Mg in MgF$_2$ and of $^{139}$La in LaF$_3$.

## Notes and references

## Graphical abstract

A fine agreement is obtained between calculated PBE-DFT and experimental $^{19}$F isotropic chemical shifts by applying an empirical correction for the description of the Ca 3d, Sc 3d and La 4f orbitals.

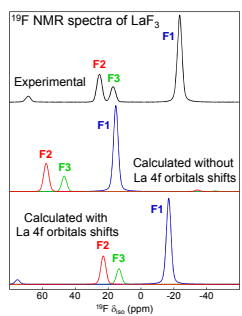

# NMR parameters in alkali, alkaline earth and rare earth fluorides from first principle calculations

Aymeric Sadoc, Monique Body, Christophe Legein, Mamata Biswal, Franck Fayon, Xavier Rocquefelte and Florent Boucher

**Electronic Supplementary Information**

**Table of content**

**Experimental conditions used for ¹⁹F solid state NMR spectroscopy**

The $^{19}$F solid-state MAS NMR experiments were conducted on Avance 300 (magnetic field of 7.0 T) and Avance 750 (magnetic field of 17.6 T) Bruker spectrometers operating at Larmor frequencies of 282.2 and 705.85 MHz, respectively, using 2.5 mm and 1.3 mm CPMAS probehead. All spectra were acquired using a Hahn echo sequence with an inter-pulse delay equal to one rotor period, except $CaF_2$ and $LaF_3$ for which a single pulse sequence was used. The recycle delays were set to 10 s for LiF, NaF, KF, RbF, CsF, $MgF_2$, $CaF_2$, $SrF_2$, $BaF_2$ and $ScF_3$ and 30 s for $YF_3$ and $LaF_3$. $^{19}$F nutation frequencies ranging between 93 (2.5 mm prohead) and 195 kHz (1.3 mm probehead) were used. The $^{19}$F chemical shifts were referenced to $CFCl_3$ at 0 ppm.

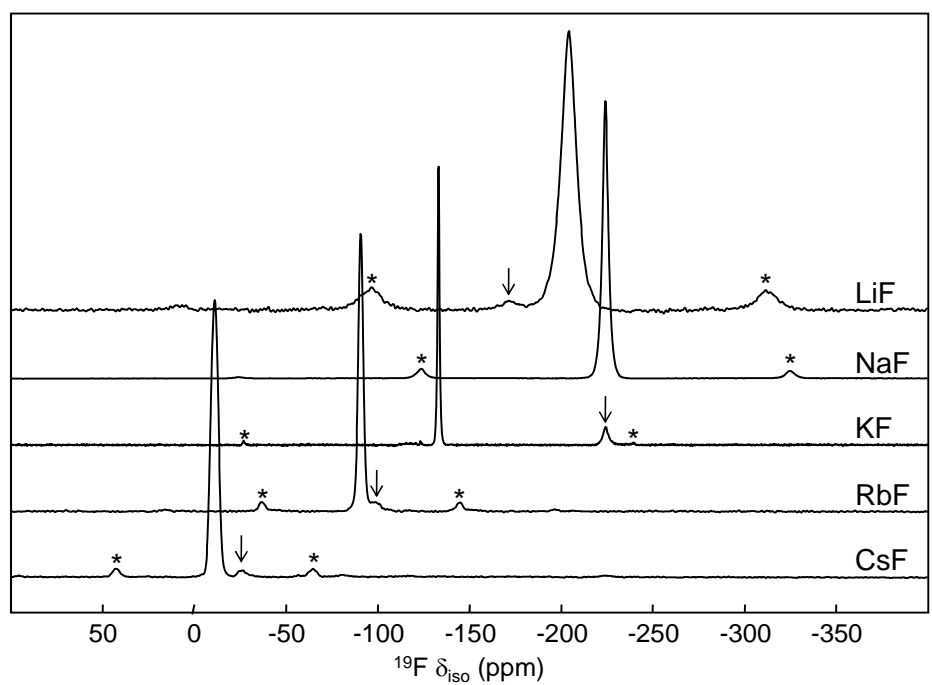

**Figure S1.** $^{19}$F MAS NMR spectra of alkaline fluorides obtained at a magnetic field of 7.0 T using spinning frequencies of 30 kHz for LiF, 25 kHz for NaF and KF and 15 kHz for RbF and CsF. The arrows on the NMR spectra of LiF, RbF and CsF indicate unidentified impurities. The arrow on the spectrum of KF indicates an impurity identified as NaF. The asterisks indicate spinning sidebands.

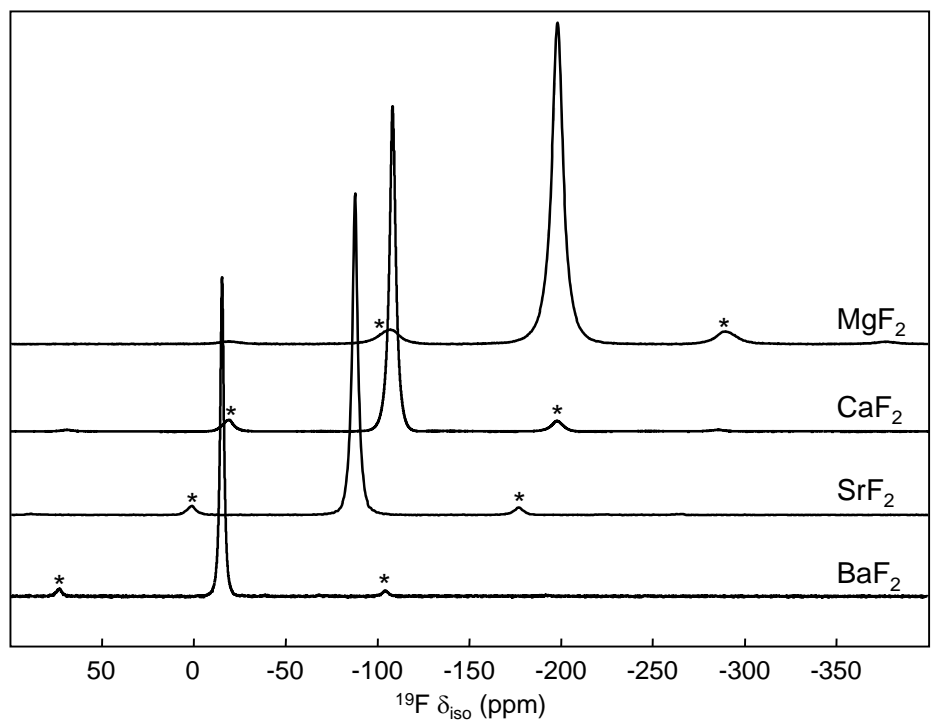

**Figure S2.** $^{19}$F MAS NMR spectra of alkaline earth fluorides obtained at a magnetic field of 7.0 T using spinning frequencies of 25 kHz. The asterisks indicate spinning sidebands.

**Solid state NMR and PXRD study of ScF$_3$**

ScF$_3$ was recently studied by both $^{19}$F and $^{45}$Sc solid-state MAS NMR but the reported results[1] appear to us somewhat surprising and the results obtained in our study were also not straightforward to interpret.

Firstly, two different crystalline structures are reported for ScF$_3$ at ambient temperature and pressure: a cubic one[2,3] (ReO$_3$ type, space group: Pm-3m) and a rhombohedral one[4-6] (distorted ReO$_3$ type, space group: R32). Lo *et al.* report that their ScF$_3$ sample adopts a rhombohedral structure. However the small 2θ range of their powder X-ray diffraction (PXRD) pattern[1] does not allow confirming this assumption since both cubic and rhombohedral structures give very similar patterns, except for large 2θ values. The PXRD patterns recorded for our sample (Aldrich, 99.99%, lot number 04937HE) on the 2θ ranges 20-125°, 117.7-119.3°, 139.5-142.5° and 146.1-149.5° are shown in Figure 3. These diagrams do not evidence any rhombohedral splitting (Table 1) indicating that ScF$_3$ adopt a cubic structure at ambient temperature and pressure, in agreement with a recent study of the pronounced negative thermal expansion (NTE) of ScF$_3$.[7]

Both cubic and rhombohedral structures of ScF$_3$ contain a single Sc site and a single F site in the unit cell. Nevertheless, we were not able to reconstruct the $^{19}$F MAS NMR spectrum with a single resonance (Figure 2). A satisfying reconstruction is obtained with three lines having close δ$_{iso}$ values but significantly different chemical shift anisotropies (Table 2). Lo *et al.* report a δ$_{iso}$ value equal to -35.9 ppm,[1] in good agreement with previously reported results,[8] and a CSA equal to ca. 305 ppm. This large value is in agreement with the observed intense spinning sidebands (the $^{19}$F reconstructed spectrum is not presented).

Moreover, Lo *et al.*[1] estimated the $^{45}$Sc quadrupolar coupling constant to 1.3(2) MHz and they claimed that "this small value is consistent with the high spherical symmetry around $^{45}$Sc" whereas, as outlined by themselves, this nucleus has a moderately sized quadrupole moment (Q = - 0.22 x 10$^{-28}$ m$^2$).[9] We have also recorded a $^{45}$Sc NMR spectrum of ScF$_3$ (Figure 3). Assuming a cubic structure, in which the Sc atom occupy the site with m-3m symmetry (1a Whyckoff position), a quadrupolar coupling constant equal to zero is expected. As Lo *et al.*[1], we observe a spinning sideband manifold, indicating quadrupolar frequency different from zero, and the shape of this spinning sideband manifolds likely indicates some disorder in the structure. This spectrum is consequently difficult to reconstruct with a single set of parameters and the quadrupolar frequency can only be roughly estimated to 20 kHz ($C_Q$=280 kHz). Whereas the determined $^{45}$Sc $\delta_{iso}$ value (-51.8 ppm) is very similar to the one determined by Lo et al. (-52 ppm), our quadrupolar coupling constant is significantly lower indicating less distorted $Sc^{3+}$ sites.

At first glance, these results which can only be explained by the presence of some structural disorder in $ScF_3$ seem puzzling. Nevertheless, disorder was previously mentioned in $ScF_3$ to explain its marked NTE.[7] The assumed mechanism, *i. e.* rocking motion of essentially rigid $ScF_6^{3-}$ octahedra, is supported by the large transverse component of the anisotropic displacement parameters (ADPs) for the fluoride anions.[7] ADPs may represent either atomic motion or static displacive disorder and static disorder was also invoked since it has been suggested for $AlF_3$ above its rhomboedral-to-cubic phase transition (the Al-F-Al links are locally bent in the cubic phase).[10,11] Both dynamic (depending on the motion frequency) and static disorders explain the non-zero quadrupolar frequency of $^{45}$Sc (Figure 3) and the several lines used for the reconstruction of the $^{19}$F NMR spectrum (Figure 2 and Table 2).

Local structural disorder in $ScF_3$ could also arise from incomplete fluorination leading to $ScF_{3-2x}O_x\square_x$ compounds and/or from occurrence of hydroxyl groups substituting fluoride ions into the network. Both these assumptions can be ruled out since the fluorinations of our sample, using either HF or $F_2$ at 600°C, do not lead to any changes on the NMR spectra.

Since several lines are used for the reconstruction of the $^{19}$F NMR spectrum, $^{19}$F $\delta_{iso}$ value for $ScF_3$ can only be roughly determined and the uncertainty is higher than for the others studied compounds. We choose the chemical shift value at the peak maximum, *i. e.* -36 ppm.

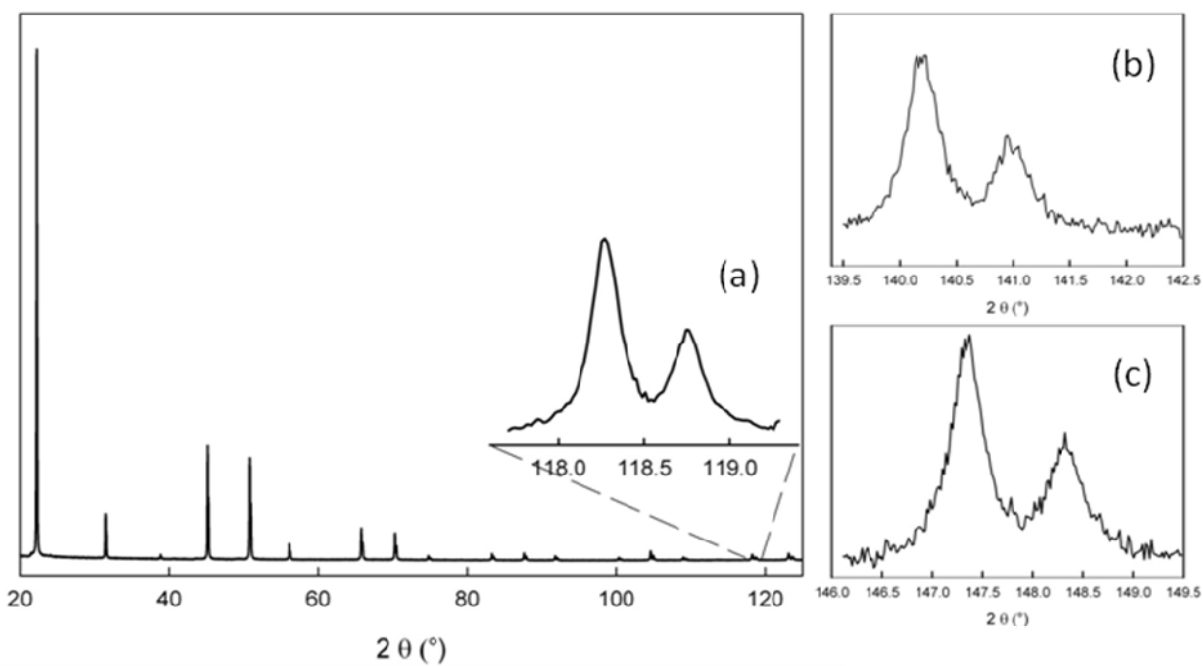

**Figure S3.** X-ray powder diffraction diagrams of $ScF_3$. In (a), (b) and (c) are shown the (4,2,0), (4,2,2) and (4,3,3) reflections, respectively. These diagrams were recorded under air, at room temperature with a PANalytical X'pert PRO diffractometer equipped with a X'Celerator detector using monochromated CuKα radiation (λ = 1.54056 Å). Measurements were done with an interpolated step of 0.017°, in the 2θ ranges 20-125°, (a) 117.7-119.3°, (b) 139.5-142.5° and (c) 146.1-149.5°, and total collecting times of 2 h 06 min, (a) 24 min, (b) 12 min and (c) 35 min.

**Table S1.** (h,k,l) reflections and corresponding 2θ values (°) of $ScF_3$ assuming Pm-3m[3] ($ICSD^{12}$ file number 36011) and R32[6] ($ICSD^{12}$ file number 77071) space groups (λ = 1.54056 Å).

| Pm-3m | | | | R32 | | | |
|---|---|---|---|---|---|---|---|
| h | k | l | 2θ | h | k | l | 2θ |
| 0 | 2 | 4 | 118.373 | 0 | 2 | 4 | 117.725 |
|   |   |   |         | -2 | 0 | 4 | 118.070 |
| 2 | 2 | 4 | 140.376 | 2 | 2 | 4 | 139.010 |
|   |   |   |         | -2 | 2 | 4 | 139.711 |
|   |   |   |         | -2 | -2 | 4 | 139.948 |
| 0 | 3 | 4 | 147.565 | 0 | 3 | 4 | 146.189 |
| 0 | 0 | 5 | 147.565 | 0 | 0 | 5 | 146.602 |
|   |   |   |         | -3 | 0 | 4 | 147.019 |

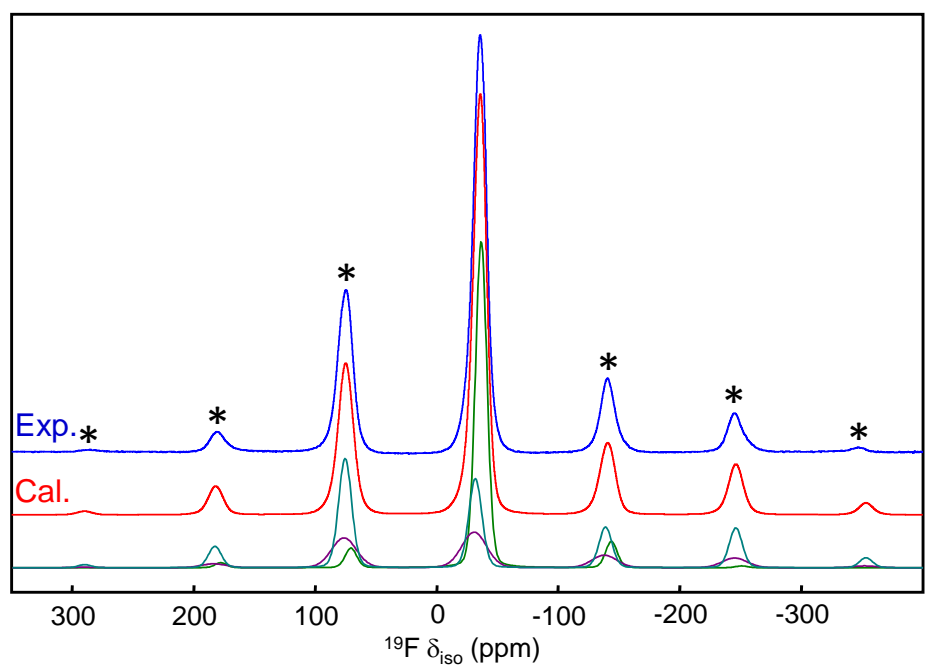

**Figure S4.** Experimental (exp.) and reconstructed (cal.) $^{19}F$ MAS NMR spectra of $ScF_3$ obtained at a magnetic field of 7.0 T using a spinning frequency of 30 kHz. The three individual contributions to the reconstructed spectrum are shown below. The asterisks indicate spinning sidebands.

**Table S2.** $^{19}F$ isotropic chemical shifts ($δ_{iso}$, ppm), chemical shift anisotropies ($δ_{aniso}$, ppm), asymmetry parameters (η), line widths and relative intensities (%) determined from the reconstruction of the $^{19}F$ NMR spectrum of $ScF_3$.

| Line | $δ_{iso}$ (±0.5) | $δ_{aniso}$ (±10) | η (±0.05) | Width (±0.5) | Intensity (±0.5) |
|---|---|---|---|---|---|
| 1 | -36.5 | 107 | 0 | 11.2 | 40.7 |
| 2 | -31.7 | -279 | 0 | 12.7 | 38.0 |
| 3 | -30.8 | -322 | 0.2 | 24.1 | 21.3 |

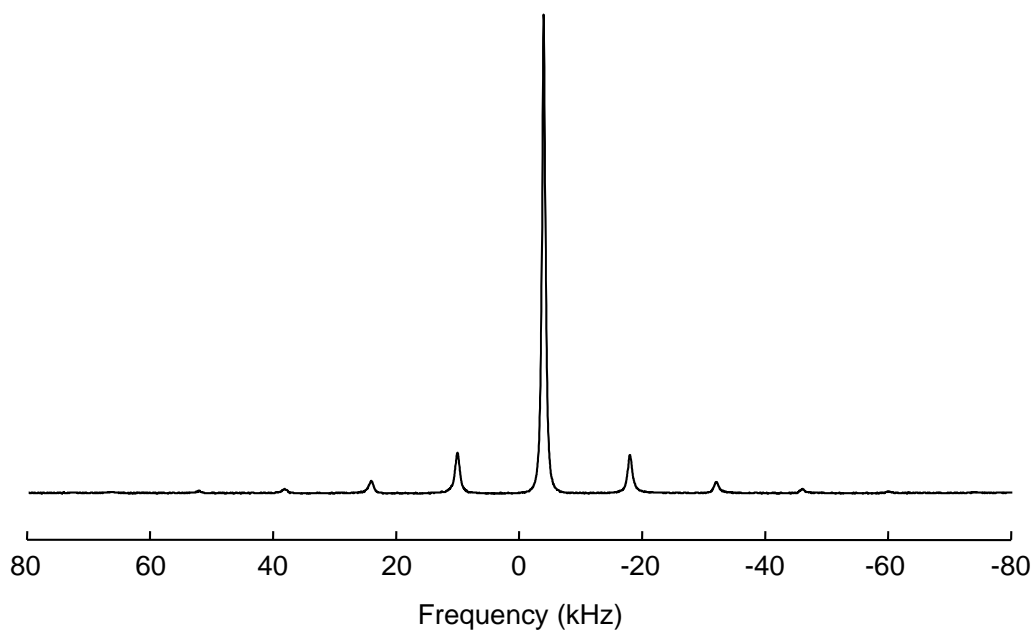

**Figure S5.** Experimental $^{45}$Sc SATRAS (SAtellite TRAnsition Spectroscopy[13,14]) MAS NMR spectrum of ScF$_3$ recorded on a Bruker Avance 300 (7.0 T) spectrometer operating at a Larmor frequency of 72.906 MHz using a 2.5 mm probehead. The spinning frequency was 14 kHz. The quantitative excitation of all transitions[15] was ensured by using a short pulse duration (1 μs) with low-radio-frequency (RF) field strength (70 kHz). The recycle delay was set to 5 s. The $^{45}$Sc chemical shift was referenced to a 1 M Sc(NO$_3$)$_3$ aqueous solution.

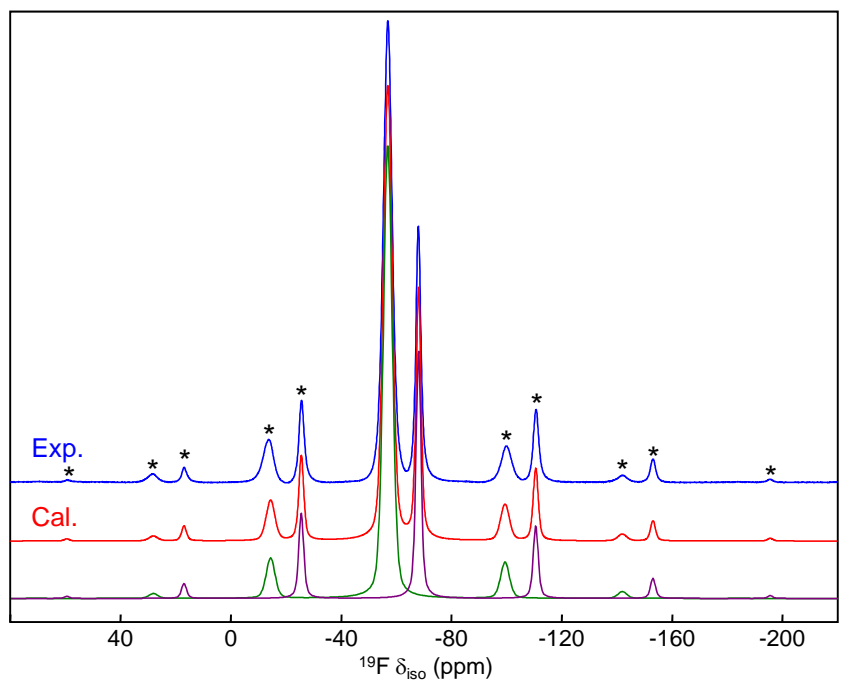

**Figure S6.** $^{19}$F experimental (exp.) and reconstructed (cal.) $^{19}$F MAS NMR spectra of YF$_3$ recorded at a magnetic field of 17.6 T using a spinning frequency of 30 kHz. The two individual contributions to the reconstructed spectrum are shown below. The asterisks indicate spinning sidebands.

**Table S3.** $^{19}$F isotropic chemical shifts ($\delta_{iso}$, ppm), chemical shift anisotropies ($\delta_{aniso}$, ppm), asymmetry parameters ($\eta$), line widths (ppm), relative intensities (%) determined from the reconstruction of the $^{19}$F NMR spectrum of YF$_3$ and line assignment.

| Line | $\delta_{iso}$ (±0.2) | $\delta_{aniso}$ (±5) | $\eta$ (±0.05) | Width (±0.1) | Intensity (±0.5) | Assignment |
|---|---|---|---|---|---|---|
| 1 | -56.9 | -42.5 | 0.75 | 3.8 | 66.4 | F2 |
| 2 | -68.1 | -76.5 | 0.80 | 2.1 | 33.6 | F1 |

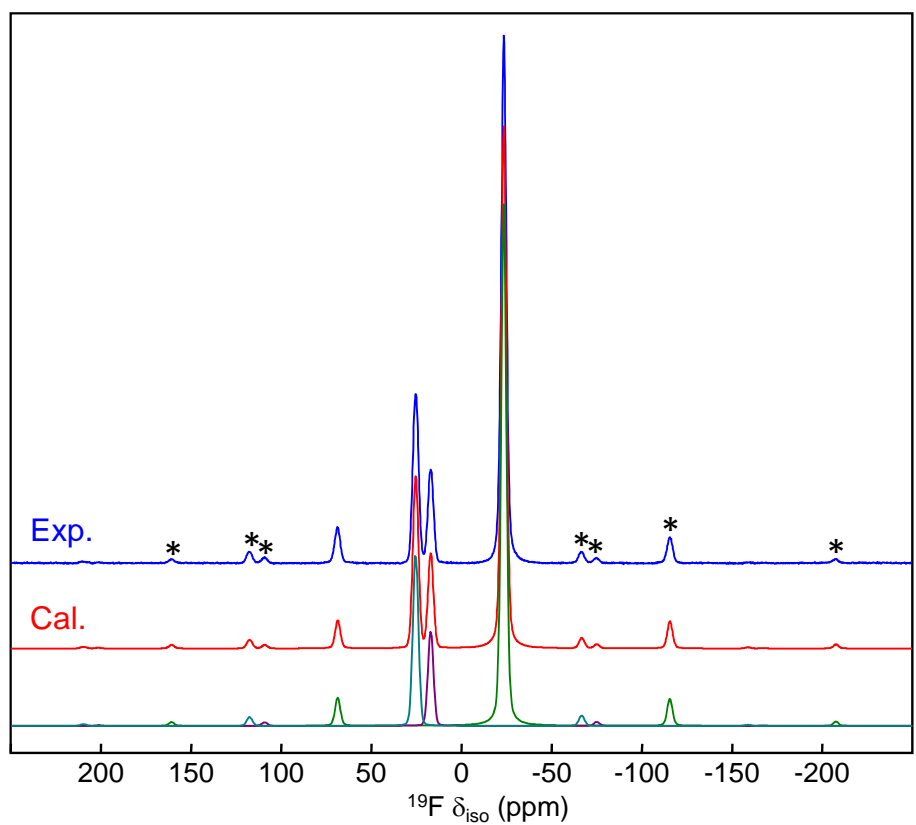

**Figure S7.** $^{19}$F experimental (exp.) and reconstructed (cal.) MAS NMR spectra of LaF$_3$ recorded at a magnetic field of 17.6 T using a spinning frequency of 65 kHz. The three individual contributions to the reconstructed spectra are shown below. The asterisks indicate spinning sidebands.

**Table S4.** $^{19}$F isotropic chemical shifts ($\delta_{iso}$, ppm), chemical shift anisotropies ($\delta_{aniso}$, ppm), asymmetry parameters ($\eta$), line widths (ppm), relative intensities (%) determined from the reconstruction of the $^{19}$F NMR spectrum of LaF$_3$ and line assignment.

| Line | $\delta_{iso}$ (±0.2) | $\delta_{aniso}$ (±5) | $\eta$ (±0.05) | Width (±0.1) | Intensity (±0.5) | Assignment |
|---|---|---|---|---|---|---|
| 1 | -23.6 | -71 | 0.9 | 3.3 | 66.7 | F1 |
| 2 | 16.9 | 66 | 0.55 | 3.5 | 11.3 | F3 |
| 3 | 25.3 | 78 | 0.55 | 3.6 | 22.0 | F2 |

**Table S5.** Fractional atomic coordinates from the initial (IS)[16] and and PBE-DFT geometry-optimized (APO) structures for $MgF_2$.

| Atom | Site | | x | y | z |
|---|---|---|---|---|---|
| Mg | 2a | | 0 | 0 | 0 |
| F | 4f | IS | *0.3028* | *0.3028* | *0* |
|   |    | APO | 0.3022 | 0.3022 | 0 |

**Table S6.** Mg-F bond lengths and F-Mg-F bond angles deduced from the initial[16] (IS) and PBE-DFT geometry-optimized (APO) structures for $MgF_2$.

| Bond lengths/Å | | Bond angles/° | |
|---|---|---|---|
| IS | APO | IS | APO |
| 1.979 | 1.975 | 81.04 | 81.22 |
| 1.984 | 1.986 | 98.96 | 98.78 |

**Table S7.** Eigenvectors of the calculated $^{25}Mg$ EFG tensor, after optimization, expressed in Cartesian coordinates for $MgF_2$ at the (0,0,0) position. The definition of the Cartesian axis with respect to the lattice parameters is given below.

| Axis | $V_{xx}$ | $V_{yy}$ | $V_{zz}$ |
|---|---|---|---|
| *i* | 0.7071 | 0 | -0.7071 |
| *j* | 0.7071 | 0 | 0.7071 |
| *k* | 0 | 1 | 0 |

With $\begin{pmatrix} a \\ b \\ c \end{pmatrix} = \begin{pmatrix} 4.6213 & 0.0000 & 0.0000 \\ 0.0000 & 4.6213 & 0.0000 \\ 0.0000 & 0.0000 & 3.0159 \end{pmatrix} \times \begin{pmatrix} i \\ j \\ k \end{pmatrix}$

and   a = b = 4.6213 Å, c = 3.0159 Å ; α = β = γ = 90°

**Table S8.** Fractional atomic coordinates from the initial (IS)[17] and PBE-DFT geometry-optimized (APO) structures for $YF_3$.

| Atom | Site | | x | y | z |
|---|---|---|---|---|---|
| Y | 4c | IS | *0.3673* | *1/4* | *0.0591* |
|   |    | APO | 0.3687 | 1/4 | 0.0604 |
| F1 | 4c | IS | *0.5227* | *1/4* | *0.5910* |
|    |    | APO | 0.5231 | 1/4 | 0.5906 |
| F2 | 8d | IS | *0.1652* | *0.0643* | *0.3755* |
|    |    | APO | 0.1655 | 0.0629 | 0.3775 |

**Table S9.** Y-F bond lengths deduced from the initial[17] (IS) and PBE-DFT geometry-optimized (APO) structures for $YF_3$.

| Bond lengths/Å | IS | APO |
|---|---|---|
| Y-F1 | 2.282 | 2.285 |
|      | 2.287 | 2.294 |
|      | 2.538 | 2.528 |
| Y-F2 | 2.281 (x2) | 2.291 (x2) |
|      | 2.299 (x2) | 2.296 (x2) |
|      | 2.310 (x2) | 2.299 (x2) |

**Table S10.** Fractional atomic coordinates from the initial (IS)[18] and PBE-DFT geometry-optimized (APO) structures for LaF$_3$.

| Atom | Site | | x | y | z |
|---|---|---|---|---|---|
| La | 6f | *IS* | *0.6598* | *0* | *1/4* |
|  |  | APO | 0.6578 | 0 | 1/4 |
| F1 | 12g | *IS* | *0.3659* | *0.0536* | *0.0813* |
|  |  | APO | 0.3688 | 0.0584 | 0.0805 |
| F2 | 4d | *IS* | *1/3* | *2/3* | *0.1830* |
|  |  | APO | 1/3 | 2/3 | 0.1825 |
| F3 | 2a |  | 0 | 0 | 1/4 |

**Table S11.** La-F bond lengths deduced from the initial[18] (IS) and PBE-DFT geometry-optimized (APO) structures for LaF$_3$.

| Bond lengths/Å | IS | APO |
|---|---|---|
| La-F1 | 2.458 (x2) | 2.457 (x2) |
|  | 2.489 (x2) | 2.477 (x2) |
|  | 2.638 (x2) | 2.629 (x2) |
|  | 3.003 (x2) | 3.038 (x2) |
| La-F2 | 2.417 (x2) | 2.415 (x2) |
| La-F3 | 2.444 | 2.458 |

**Table S12.** Eigenvectors of the calculated $^{139}$La EFG tensor in LaF$_3$ for IS, expressed in a Cartesian coordinate system (*i, j, k*) and along the crystallographic axis (*a, b, c*) for the La position at (0.6578, 0, 1/4). The definition of the Cartesian axis with respect to the lattice parameters is given below.

| Axis | $V_{xx}$ | $V_{yy}$ | $V_{zz}$ |
|---|---|---|---|
| *i* | 0.2880 | -0.8660 | -0.4087 |
| *j* | 0.4989 | 0.5 | -0.7079 |
| *k* | 0.8174 | 0 | 0.5761 |
| *a* | 0.0466 | -0.1391 | -0.0654 |
| *b* | 0.0933 | 0.0000 | -0.1308 |
| *c* | 0.1106 | 0.0000 | 0.0789 |

With
$$\begin{pmatrix} a \\ b \\ c \end{pmatrix} = \begin{pmatrix} 6.2224 & -3.5925 & 0 \\ 0 & 7.1850 & 0 \\ 0 & 0 & 7.3510 \end{pmatrix} \times \begin{pmatrix} i \\ j \\ k \end{pmatrix}$$

and   a = b = 7.1850 Å, c = 7.3510 Å ; α = β = 90°, γ = 120°

**Table S13.** Experimental $^{19}$F isotropic chemical shifts relative to CFCl$_3$ and calculated $^{19}$F isotropic shieldings for the eleven compounds additionally considered in Figure 5. The "calculated" $^{19}$F isotropic chemical shifts according to $\delta_{iso}$/CFCl$_3$ = -0.80(3) $\sigma_{iso}$ + 89(9) are also reported.

| Compounds | Site | $\sigma_{iso}^{calc.}$ (ppm) | $\delta_{iso}^{calc.}$ (ppm) | $\delta_{iso}^{exp.}$ (ppm) |
|---|---|---|---|---|
| CdF$_2$ | F1 | 350.9[a] | -192 | -190.7[c] |
| HgF$_2$ | F1 | 356.4[a] | -196 | -197.6[d] |
| α-PbF$_2$ | F1 | 140.8[a] | -24 | -20.5[e] |
|  | F2 | 176.6[a] | -52 | -57.7[e] |
| α-AlF$_3$ | F1 | 336.2[a] | -180 | -172.0[f] |
| Na$_5$Al$_3$F$_{14}$ | F1 | 358.9[a] | -198 | -191.4[g] |
|  | F2 | 326.1[a] | -172 | -165.0[g] |
|  | F3 | 356.9[a] | -197 | -189.5[g] |
| ZnF$_2$ | F1 | 363.0[b] | -201 | -200.7[c] |
| GaF$_3$ | F1 | 314.0[b] | -162 | -167.2[c] |
| InF$_3$ | F1 | 364.1[b] | -202 | -206.2[c] |
| BaLiF$_3$ | F1 | 238.8[b] | -102 | -98.2[c] |
| β-BaAlF$_5$ | F1 | 307.4[b] | -157 | -154.6[h] |
|  | F2 | 287.9[b] | -141 | -138.9[h] |
|  | F3 | 268.5[b] | -126 | -121.3[h] |
|  | F4 | 254.9[b] | -115 | -109.2[h] |
|  | F5 | 302.4[b] | -153 | -148.8[h] |
|  | F6 | 277.5[b] | -133 | -127.5[h] |
|  | F7 | 293.3[b] | -146 | -140.8[h] |
|  | F8 | 245.4[b] | -107 | -99.0[h] |
|  | F9 | 271.7[b] | -128 | -124.5[h] |
|  | F10 | 297.0[b] | -149 | -144.6[h] |
| Ba$_3$Al$_2$F$_{12}$ | F1 | 310.9[b] | -160 | -153.3[h] |
|  | F2 | 308.2[b] | -158 | -151.6[h] |
|  | F3 | 165.1[b] | -43 | -30.5[h] |
|  | F4 | 186.6[b] | -60 | -50.8[h] |
|  | F5 | 267.6[b] | -125 | -115.7[h] |
|  | F6 | 265.4[b] | -123 | -113.0[h] |
|  | F7 | 279.8[b] | -135 | -127.9[h] |
|  | F8 | 302.3[b] | -153 | -146.4[h] |

[a] Calculated values from reference [19].
[b] Calculated values from reference [20].
[c] Experimental values from reference [21].
[d] Experimental values from reference [22].
[e] Experimental values from reference [23].
[f] Experimental values from reference [24].
[g] Experimental values from reference [25].
[h] Experimental values from reference [26].